%% file: aiml.tex
\documentclass[twoside]{aiml}

\usepackage{aimlmacro}

\usepackage{graphicx}
\usepackage{amsmath,amssymb,amsfonts}
\usepackage{graphicx}
\usepackage{tabularx,array,xcolor,multirow}
\usepackage{arydshln}
\usepackage{array}

\usepackage{tikz}
\usetikzlibrary {shapes.multipart}
\usepackage{siunitx}

\usepackage{subcaption}
\usepackage{graphicx}

\renewcommand{\phi}{\varphi}
\newcommand{\agt}{\mathcal{I}}

\usepackage{mathabx}

\usepackage{comment}

\usepackage{stackengine}

\usepackage{stmaryrd} 

\def\lastname{Dong}

\begin{document}

\begin{frontmatter}
  \title{Permissible Knowledge Pooling}
  \author{Huimin Dong}
  \address{University of Luxembourg \\ 2, place de l'Universit\'{e}\\ huimin.dong@uni.lu}

  \begin{abstract}
  Information pooling has been extensively formalised across various logical frameworks in distributed systems~\cite{aagotnes2017resolving,van2022pooling,castaneda2023communication,baltag2023learning}, characterized by diverse information-sharing patterns. These approaches generally adopt an intersection perspective, aggregating all possible information, regardless of whether it is known or unknown to the agents. In contrast, this work adopts a unique stance, emphasising that sharing knowledge means distributing what is known, rather than what remains uncertain. This paper introduces new modal logics for knowledge pooling and sharing, ranging from a novel language of knowledge pooling to a dynamic mechanism for knowledge sharing. It also outlines their axiomatizations and discusses a potential framework for permissible knowledge pooling.
  \end{abstract}

 \end{frontmatter}

\section{Motivation}
What disclosures are considered sensitive information leaks? What messages can be shared with a cooperative partner without breaching confidentiality agreements? In the processes of compliance checking, legal decision-making, and responsibility allocation, it is necessary not only to grasp the ethical ramifications but also to understand the types of epistemic information at play and their interaction with ethical considerations.

A notable gap exists in our understanding of the nature of epistemic information produced by these ethically questionable actions and the manner in which this information can shed light on the suitability or unsuitability of such actions. To address the above questions, existing research tends to focus either on the mechanics of information transfer inherent in such communicative acts~\cite{van2007dynamic,baltag2008qualitative,van2011logical,van2012logic}, or on the ethical considerations presumed by such actions~\cite{broersen2001boid,broersen2004action}. However, these approaches often overlook a comprehensive perspective. 

The mishandling of sensitive information~\cite{aucher2010privacy}, especially in the context of multi-agent interactions, exemplifies the complexities involved in addressing such issues. This challenge underscores the need for a more integrated approach that considers both the mechanics of information exchange and the ethical implications thereof.

\begin{example}[Sensitive Information]
A customer purchases services from a company to analyse user behaviors on websites. The company offers two servers, $a$ and $b$, each providing different type of information. Server $a$ tracks how users ($p$) are associated with user IDs ($q$), while server $b$ details how these user IDs correspond to the websites they visit ($r$). While the company is \emph{able} to sell access to both servers, GDPR regulations \emph{prohibit} selling both simultaneously, as doing so would result in a breach of privacy by revealing users' personal website visitation patterns to the customer.  

The knowledge provided by the servers can be represented as follows:
\begin{itemize}
    \item $K_a(p \to q)$ and
    \item $K_b (q \to r)$.
\end{itemize}
GDPR restricts the customer from acquiring knowledge that links users directly to the websites they visit:
\begin{itemize}
    \item $K_c(p \to r)$.
\end{itemize}
Given that the business transaction requires at least one server to relay information to the customer, how can we construct a model of \emph{permissible} knowledge transfer to investigate the underlying mechanisms?
\end{example}

\begin{center}
    {\bf Research Question:} How do we determine the ethical considerations in sharing information? 
\end{center}
This research question unfolds into two distinct inquiries:
\begin{enumerate}
    \item What mechanisms underpin the act of sharing information? 
    \item How do the epistemic and ethical implications of communication converge to define the permissibility of such actions? 
\end{enumerate}

This work will address these questions as outlined below. Section~\ref{sec:share} will introduce a dynamic model of knowledge sharing, highlighting the role of interpresonal dependence in the exchange of information. Following the tradition of dynamic deontic logic~\cite{van1996dynamic}, Section~\ref{sec:perm} will develop a new framework to differentiate between the permissions to know and to share information. Section~\ref{sec:tech} will discuss various results concerning expressivity, axiomatization, and completeness. Section~\ref{sec:related} will review relevant literature and Section~\ref{sec:con} will provide concluding remarks.

\section{Theory of Knowledge} \label{sec:share}

\subsection{Static Information}

We follow the standard approach of distributed knowledge~\cite{van2007dynamic} to examine the \emph{static} dimension of information distribution and sharing. Within this framework, the modality $K_a$ represents the individual knowledge of agent $a$, while the modality $D_G$ signifies the distributed knowledge within a group $G$. This allows us to express that ``Agent $a$ knows $\varphi$'' using $K_a\varphi$ and that ``Group $G$ (distributively) knows $\varphi$ if every members shares their knowledge'' with $D_G\varphi$. 

An agent's knowledge can be augmented through the acquisition of insights from another. For instance, a customer's understanding of user behaviours on websites is reinforced when one server shares its knowledge. This \emph{agent-dependent} knowledge will be expressed by a new modality $K^a_b$. The expression $K^a_b\varphi$ represents that agent $b$ possesses knowledge of $\varphi$ dependent on the input from agent $a$. 
\begin{definition}[Language]
The language $\mathcal{L}$ of \emph{agent-dependent} knowledge is defined as follows:
\begin{align*}
    \phi:= p \mid \neg\phi \mid \phi \wedge \phi \mid K_a\phi \mid D_G\phi \mid K^a_b\varphi
\end{align*}
where $p \in Prop$ is an element of the (countable) set $Prop$ of atomic propositions, $a \in \mathcal{I}$ is an element of the (finite) set $\mathcal{I}$ of agents, and $G \subseteq \mathcal{I}$ is a subset of agents. 
\end{definition}   
As usual, the dual modality $\hat{K_a}\phi$ of individual knowledge is defined as $\neg K_a \neg\phi$, which expresses the possibility of knowing individually. The ``everybody knows'' knowledge for group $G$, denoted as $E_G\varphi$, can be defined as $\bigwedge_{a \in G}K_a\varphi$. 

\begin{definition}[Models]
A structure $M = \langle W, \{R_a\}_{a \in \mathcal{I}}, V\rangle$ is a model when:
\begin{itemize}
    \item $W$ is a non-empty set of possible states;
    \item $R_a \subseteq W \times W$ is an equivalence relation;
    \item $V: Prop \to 2^{W}$ is a valuation function. 
\end{itemize}
\end{definition}

\noindent
The truth conditions of Boolean formulas and knowledge modalities are defined as usual~\cite{van2007dynamic}:
\begin{align*}
    M, w \models K_a \phi & \textrm{ iff } R_a[w] \subseteq ||\phi||_M;\\
    M, w \models D_G \phi & \textrm{ iff } D_G[w] \subseteq ||\phi||_M,
\end{align*}
where $D_G = \bigcap_{a \in G}R_a$, $R_a[w] = \{ u\in W \mid (w, u) \in R_a\}$, and $||\phi||_M = \{w \in W \mid M, w \models \phi\}$. When the model $M$ is clear from the context, we simplify the notation $||\phi||_M$ to $||\phi||$. We define $\equiv_{a:w} \subseteq W\times W$ as an equivalence relation based on agent $a$'s basic knowledge at state $w$: $s \equiv_{a:w} u$ if and only if $s \in ||\varphi|| \Leftrightarrow u \in ||\varphi||$ for all $\varphi \in K_aw$, where $K_aw=\{\varphi \in\mathcal{L}\mid M, w \models K_a\varphi\}$ is agent $a$'s knowledge at $w$. Now the agent-dependent knowledge is interpreted as follows:
\[
M, w\models K^a_b\varphi \text{ iff } (R_b \cap \equiv_{a:w})[w]\subseteq ||\varphi||. 
\]
The relation $R_b \cap \equiv_{a:w}$ is an equivalence relation. 

The expression $K^a_b\varphi$ analyses $b$'s knowledge by integrating the relevant knowledge of $a$. This integration is interpreted by the \emph{intersection} between $b$'s knowledge accessible relation $R_b$ and a corresponding categorization of agent $a$'s knowledge. Essentially, it can be understood as ``Agent $b$ knows $\varphi$ when considering the knowledge from $a$ remain constant.'' This approach of knowledge, grounded in equality, reflects the concept of \emph{ceteris paribus} (with all other things else being equal)~\cite{van2009everything,grossi2013ceteris,girard2016ceteris} in the perspective of epistemology.  

These static modalities of knowledge are considered standard. The modality of individual knowledge $K_a$ and the modality of distributed knowledge are both characterized by the {\sf S5} axiomatic frameworks~\cite{blackburn2001modal}. The agent-dependent modality $K^a_b$ is an {\sf S5} modality as well therefore, capturing the essence of knowledge. However, the interactions between them extend the standard {\sf S5} frameworks. 
\begin{itemize}
    \item ({\sf Int}) $K_a\varphi \to K^b_a\varphi$
    \item ({\sf Cl}) $K^b_a\varphi\to\bigvee_{\psi \in \mathcal{L}}(K_a\psi \wedge K_b(\psi\to\varphi))$
\end{itemize}
Details about axiomatizations of these modalities are presented in Section~\ref{sec:ak}.


The following logical implication captures the relationship among these three types of knowledge: 
$$K_b\varphi \to K^a_b\varphi \to D_{\{a,b\}}\varphi.$$
Agent-dependent knowledge serves as a bridge from individual to distributed knowledge, filling the connection from individual towards distributed knowledge. The two examples below further clarify this relationship.

\begin{example}[Sensitive Information, Cont'd]
\label{exm:nd}
Figure~\ref{fig:RG} illustrates a model $M = \langle W, R_a, R_b, R_c, V\rangle$ representing the individual knowledge of agents $a$ and $b$, where both agents only know $p$:
\begin{itemize}
    \item $W = \{s_0,s_1,s_2,s_3,s_4\}$,
    \item $R_a = \{(s,s) \mid s\in W\} \cup \{(s_0,s_1),(s_1,s_0), (s_0,s_2),(s_2,s_0)\}$, 
    \item $R_b = \{(s,s) \mid s\in W\} \cup \{(s_0,s_3),(s_3,s_0),(s_0,s_4),(s_4,s_0)\}$,
    \item $R_c = R_a \cup R_b$
    \item $V(p) = \{s_0,s_1,s_3\}$, $V(q) = \{s_0,s_1,s_2\}$, and $V(r) = \{s_0,s_2\}$.
\end{itemize}
At $s_0$, the following individual knowledge about $a$, $b$, and $c$ are satisfied:
\begin{itemize}
    \item $K_a(p\to q), \neg K_a(q \to r), \neg K_a(p \to r)$;
    \item $\neg K_b(p\to q), K_b(q \to r), \neg K_b(p \to r)$;
    \item $\neg K_c(p\to q), \neg K_c(q \to r), \neg K_c(p \to r)$.
\end{itemize}

During knowledge pooling, the flow of information from server $a$ to customer $c$ or vice versa is crucial for generating distinct agent-dependent knowledge outcomes. Agent $c$'s knowledge can be augmented based on the classification from $a$, while $a$'s knowledge dependent on $c$ remains the same:  
\begin{itemize}
    \item $K^c_a(p\to q), \neg K^c_a(q \to r), \neg K^c_a(p \to r)$;
    \item $K^a_c(p\to q), \neg K^a_c(q \to r), \neg K^a_c(p \to r)$.
\end{itemize}
The further information pooling from server $b$ to $c$ leads to:
\begin{itemize}
    \item $K_c^{ab}(p\to q), K_c^{ab}(q \to r), K_c^{ab}(p \to r)$.~\footnote{The truth condition of $K_c^{ab}$ can be roughly understood as:
    $$M,w \models K_c^{ab} \text{ iff } (R_c\cap \equiv_{a:w} \cap \equiv_{b:w})[w]\subseteq||\varphi||.$$}
\end{itemize}
Customer $c$ now knows the full information the rules of user behaviors. However, it does not imply that the customer $c$ is supposed to know the dataset of the users:
\begin{itemize}
    \item $\neg K_c^{ab}p$.
\end{itemize}

In contrast, distributed knowledge generated through information pooling does not merely copy the individual agent-based insights; it produces knowledge that extends well beyond the possible knowable limit of all individuals' understanding. 
\begin{itemize}
    \item $D_{\{a,b,c\}}(p \to q), D_{\{a,b,c\}}(q\to r), D_{\{a,b,c\}}(p \to r)$;
    \item $D_{\{a,b,c\}}p, D_{\{a,b,c\}}q,D_{\{a,b,c\}}r$.
\end{itemize}
    \begin{figure}[h!]
    \centering
    \input{TabFig/fig_3k.tex}
    \caption{Knowledge sharing: Beyond the aggregation of arbitrary information for distributed knowledge.}
    \label{fig:RG}
\end{figure}
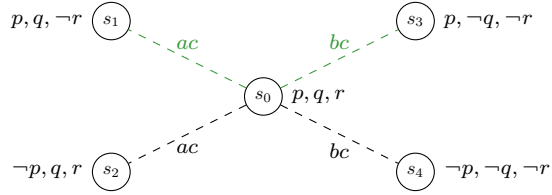
\end{example}
Essentially, agents only share what they have known. For example, agent $a$ shares her knowledge of $p \to q$ with $c$, which $c$ does not know. They cautiously avoid extending to what they are uncertain. Neither $a$ nor $c$ can verify or falsify $p$ nor $q$; this information is still uncertain even after they share knowledge with each other. Distributed knowledge represents a special boundary of information pooling, characterized by the aggregation of all conceivable information into collective knowledge, even if gathering those uncertain goes beyond what can be verified. Conversely, knowledge pooling is specifically aimed at gathering confirmed or known information between agents.


A primary concern in knowledge pooling is about communication. The expression $K_a^b(p \wedge \neg K_a p)$, a variant of Moore sentence~\cite{holliday2010moorean,hintikka1962knowledge}, remains logically consistent. However, the interpretation of ``$p$ is true but agent $a$ does not know $p$'' becomes problematic because it leads to unsuccessful knowledge sharing, especially after agents have fully communicated with each other~\cite{holliday2010moorean,fitch1963logical}. A potential solution to this challenge is to introduce a \emph{dynamic} mechanism for knowledge sharing. After knowledge sharing, $K_a^b(p \wedge \neg K_a p)$ will become inconsistent, while $K_a^b(p \wedge K_a p)$, which indicates a successful delivery of knowledge, will becomes true. This effective communication of knowledge pooling is crucial in evaluating the ethical and normative implication of such interaction. These dynamic processes will be explored in the next Section~\ref{sec:dyn} and their ethical considerations will be discussed in Section~\ref{sec:perm}.

\subsection{Knowledge Sharing: A New Type of Information Pooling} \label{sec:dyn}

Information flows from one agent to another, called information pooling, is a \emph{dynamic} process. It typically involves aggregating all available information from the participating agents through information intersection~\cite{aagotnes2017resolving,van2022pooling,castaneda2023communication,baltag2023learning}, without necessarily confirming whether the information constitutes knowledge. To ensure the accuracy and integrity of communication, this section posits that information pooling ought to involve the sharing of knowledge. To express the concept of knowledge sharing between two agents, where agent $b$ receives agent $a$'s knowledge, we introduce the notation $a \smalltriangleright b$ to denote this dynamic communication. Accordingly, the symbol $[a \smalltriangleright b]$ represents the act of sender $a$ conveying its knowledge to receiver $b$. Now we add the formula $[a \smalltriangleright b]\phi$ into the language and interpret this formula as follows: 
\begin{definition}[Updated Model] \label{def:update}
Let $M = \langle W,  \{R_a\}_{a \in \mathcal{I}}, V\rangle$ be a model and $w \in W$ be a point in $M$. Then:
\[
M, w \models [a \smalltriangleright b]\phi \textrm{ iff } M|_{a \smalltriangleright b}, w \models \phi
\]
where $M|_{a \smalltriangleright b} = \langle W,  \{R^*_i\}_{i \in \mathcal{I}}, V\rangle$ is a $(a \smalltriangleright b)$-updated model of knowledge sharing such that
\begin{enumerate}
    \item If $i \neq b$, then $R^*_i = R_i$; \label{def:u1}
    \item \label{def:u2} If $i = b$, then $(s,u)\in R_b^*$
    \begin{enumerate}
    \item When $s,u\in R_b[w]$, $(s,u)\in R_b \cap \equiv_{a:w}$;
    \item When $s,u\not\in R_b[w]$, $(s,u)\in R_b$.
    \end{enumerate}
    
\end{enumerate} 
\end{definition}
The primary feature of our updated mechanism is its \emph{directional} nature. It only eliminates those epistemic relations accessible to the recipient that do not align with the sender's knowledge. This approach assumes that the recipient's knowledge is dependent on the knowledge shared by the sender. 

In addition, this updating process is \emph{syntactic}. The fundamental principle of this knowledge sharing update is represented by intersecting $b$'s knowledge $R_b$ with the equivalence set $\equiv_{a:w}$ which reflects the sender $a$'s knowledge locally at $w$. As a result, the model is modified according to syntactic expressions that represent the sender's knowledge, ensuring that the receiver adjusts their knowledge while maintaining \emph{every thing else being equally characterising the sender's knowledge}, a variant of the concept known as ``\emph{ceteris paribus}''~\cite{van2009everything,grossi2013ceteris,girard2016ceteris}. This maintains the integrity of the sender's knowledge in the recipient's understanding.  

\begin{example}[Sensitive Information, Cont'd]
Revisiting the model $M$ illustrated in Figure~\ref{fig:RG}, we present some findings on Moore sentences that illuminate the differences between static and dynamic knowledge arising from knowledge sharing:
\begin{itemize}
    \item $K^a_c((p \to q) \wedge \neg K_c(p \to q))$ is true at $w$ in $M$;
    \item $K^a_c((p \to q) \wedge \neg K_c(p \to q))$ is false at $w$ in $M|_{a\smalltriangleright b}$;
    \item $K^a_c((p \to q) \wedge  K_c(p \to q))$ is true at $w$ in $M|_{a\smalltriangleright b}$.
\end{itemize}    
\end{example}



The update process of knowledge sharing implies two key points. Firstly, it categorizes the recipient's knowledge based on the knowledge provided by the sender, as shown in Lemma~\ref{lem:update}.\ref{lem:u4}. Secondly, the modification of agent-dependent knowledge does not influence the distributed knowledge, as demonstrated in Lemma~\ref{lem:update}.\ref{lem:u5}. 
\begin{lemma} \label{lem:update}
Let $M|_{a \smalltriangleright b} = \langle W,  \{R^*_i\}_{i \in \mathcal{I}}, V\rangle$ be a $(a \smalltriangleright b)$-updated model. 
\begin{enumerate}
     \item $M|_{a \smalltriangleright b}$ is still a model.  \label{lem:u2}
    \item \label{lem:u4}$R_a \cap R_b \subseteq R^*_b$.
    \item \label{lem:u5}$D_G = D_G^*$.
\end{enumerate}
\end{lemma}
\begin{proof}
\begin{enumerate}
    \item We only need to demonstrate that $R^*_b$ is still an equivalence relation. This requires $\equiv_a$ to be an equivalence relation, which is not difficult to be proven given Definition~\ref{def:update}. 
    \item We only need to show: $R_a \subseteq \equiv_a$. Let $uR_as$ and $\varphi\in \mathcal{L}$.
    \begin{itemize}
        \item Let $M, u \models K_a\varphi$. Due to the 4 axiom of $K_a$, we have $M, u \models K_aK_a\varphi$. Since $uR_as$, so $M, s \models K_a\varphi$.
        \item Let $M, u \not\models K_a\varphi$. Due to the 5 axiom of $K_a$, we have $M, u \models K_a\neg K_a\varphi$. Since $uR_as$, so $M, s \models \neg K_a\varphi$, which is $M, s \not\models K_a\varphi$.
    \end{itemize}
    \item We know $D_G^* = R^*_b \cap D_{G\setminus\{b\}}$. Due to $R^*_b = R_b \cap \equiv_a$, so $R^*_b \subseteq R_b$. Thus, $D_G^* \subseteq D_G$. On the other hand, due to Lemma~\ref{lem:update}.\ref{lem:u4}, $R_a \cap R_b \subseteq R^*_b$, so $R_a \cap R_b \cap D_{G\setminus\{b\}} \subseteq R^*_b \cap D_{G\setminus\{b\}}$. This is $D_G \subseteq D^*_G$. 
\end{enumerate}
\end{proof}

In this work, we adopt a non-reductionist approach~\cite{wang2013axiomatizations} to characterise the properties of dynamic knowledge sharing. Traditionally, the reduction approach~\cite{van2007dynamic} has been the predominant method for modelling updates in knowledge. This typically involves a reduction axiom, specifically the axiom of dynamic iteration, which aims to simplify the iteration $[a\smalltriangleright b][c\smalltriangleright d]$ into a less complex format. However, such simplification is not feasible within our framework. Consequently, we explore an alternative non-reductionist approach. 

\begin{proposition}
    These properties are valid in our updated model. 
\begin{itemize}
    \item ({\sf Inv}) $(p \to [a\smalltriangleright b]p) \wedge (\neg p \to [a\smalltriangleright b]\neg p)$
    \item ({\sf Rev}) $\neg[a\smalltriangleright b]\varphi \to [a\smalltriangleright b]\neg\varphi$
    \item ({\sf D}) $[a\smalltriangleright b]\neg\varphi \to \neg[a\smalltriangleright b]\varphi$ 
    \item ({\sf Int$^+$}) $[a\smalltriangleright b]K_b\varphi \to \bigvee_{\psi\in\mathcal{L}}(K_b[a\smalltriangleright b](\psi \to \varphi) \wedge K_a[a\smalltriangleright b]\psi )$
    \item ({\sf Int$_+$}) $K_b[a\smalltriangleright b]\varphi \to [a\smalltriangleright b]K_b\varphi$
    \item ({\sf Int$^-$}) $[a\smalltriangleright b]K_c\varphi \leftrightarrow K_c[a\smalltriangleright b]\varphi$ where $b \neq c$
    \item ({\sf K$_{\smalltriangleright}$}) $[a\smalltriangleright b](\varphi \to \psi) \to ([a \smalltriangleright b]\varphi \to [a\smalltriangleright b]\psi)$ 
    \item ({\sf N$_s$}) From $\varphi$ infer $[a\smalltriangleright b]\varphi$
    \item ({\sf Inc$_{\smalltriangleright}$}) From $\varphi \to [a\smalltriangleright b]\psi$ infer $K_a\varphi \to [a\smalltriangleright b]K_b\psi$
\end{itemize}
\end{proposition}
These properties establish a sound and complete logical system for dynamic knowledge sharing (detailed in Section~\ref{sec:iks}). The {\sf Inv} axiom ensures that truth assignments remain invariant in the models before and after an update. The {\sf Rev} axiom reflects that the $(a\smalltriangleright b)$-sharing update always results in a unique model. The {\sf D} axiom is crucial, illustrating the feasibility of $(a\smalltriangleright b)$-sharing. The {\sf Int$^+$, Int$_+$, Int$^-$} axioms, and the {\sf Inc$_{\smalltriangleright}$} rule illustrate the interactions between the knowledge and sharing. Finally, the {\sf K$_{\smalltriangleright}$} axiom and {\sf N$_s$} rule are considered standard. These properties infer the following two standard results.

\begin{proposition}
These properties are valid:
\begin{itemize}
 \item {\sf C}: $[a\smalltriangleright b]\varphi \wedge [a\smalltriangleright b]\psi \to [a\smalltriangleright b](\varphi \wedge \psi)$
\item {\sf RK}: From $\vdash \varphi \wedge \varphi '\to \psi$, it infers $\vdash [a\smalltriangleright b]\varphi \wedge [a\smalltriangleright b]\varphi' \to [a\smalltriangleright b]\psi$.
\item {\sf RM$_{\smalltriangleright}$}: From $\varphi_1,\dots, \varphi_n \to [a\smalltriangleright b]\psi$ inferring $K_c\varphi_1,\dots, K_c\varphi_n \to [a\smalltriangleright b]K_c\psi$.
\end{itemize}
\end{proposition}

Using a fragment of our language, we can illustrate how one can transfer and share one's knowledge with others. 
\begin{proposition}
The following properties are valid for the fragment regarding $Prop^+$, the set of Boolean propositions that are only constructed from atomic propositions, negation, and conjunction. 
\begin{itemize}
 \item ({\sf Boolean}) $\varphi \leftrightarrow [a\smalltriangleright b]\varphi$ where $\varphi \in Prop^+$
 \item ({\sf Remain}) $K_c\varphi \to [a\smalltriangleright b]K_c\varphi$ where $\varphi \in Prop^+$
 \item ({\sf Sharing}) $K_a\varphi \to [a\smalltriangleright b]K_b\varphi$ where $\varphi \in Prop^+$
 \item ({\sf Step}) $[a\smalltriangleright b]K_b\varphi \to [a\smalltriangleright b][b\smalltriangleright c]K_c\varphi$ where $\varphi \in Prop^+$
 \item ({\sf Dist}) $[a\smalltriangleright b] K_b\varphi \to D_{\{a,b\}}[a\smalltriangleright b]\varphi$
\end{itemize}
\end{proposition}


Axioms {\sf Boolean}, {\sf Remain}, and {\sf Sharing} explain the process through which the knowledge of agent $a$ is accurately conveyed and integrated into the knowledge of another agent. Thus, the knowledge pooling from agent $a$ to agent $b$ augments the knowledge of the recipient $b$, without similar influence on the sender $a$ nor any third party such as agent $c$. This represents a \emph{unidirectional} flow of information. The {\sf Dist} axiom further presents the unidirectional flow of knowledge pooling can be transferred through multi-agent interactions. 

Axiom {\sf Dist} indicates that the knowledge derived from the pooling process contributes to the collective understanding shared from the sender to the recipient. However, it is not implied that all distributed knowledge can be generated from the knowledge pooling process. This distinction is clarified in Example~\ref{exm:res} presented below. Through this, knowledge pooling reaches a new type of collective knowledge that bridges individual knowledge and distributed knowledge.

In this multi-agent context, our goal is to illustrate that the resolution of knowledge within the knowledge pooling process is not merely a matter of aggregating information arbitrarily. Consider a model $M = \langle W,  \{R_a\}_{a \in \mathcal{I}}, V\rangle$. The \emph{information} resolution operator $R^i_G\phi$, which represents the pooling of arbitrary information, is defined as per existing literature~\cite{aagotnes2017resolving}:
\[
M, w \models R^i_G\phi \textrm{ iff } M|_G, w \models \phi,
\]
where $M|_G = \langle W, \{R_a|_G\}_{a \in \mathcal{I}}, V\rangle$ such that 
\begin{equation*}
R_a|_G = \begin{cases}
\bigcap_{b\in G} R_b & \quad a \in G, \\
R_a & \quad \textrm{otherwise}.
\end{cases}
\end{equation*}
While, let $G = \{a_1,\dots, a_n\}$, we can define the notion of \emph{knowledge resolution} by our operator of \emph{collective} knowledge pooling: 
\[
R_{G}\varphi:=[a_1\smalltriangleright a_2]\dots[a_{n-1}\smalltriangleright a_n][a_n\smalltriangleright a_{n-1}]\dots[a_2\smalltriangleright a_1]\varphi.
\]
The following statement illustrates a crucial distinction from knowledge to information pooling:
\begin{itemize}
    \item ({\sf Int$^{R}$}) $R_GK_a\varphi  \to R^i_GK_a\varphi$.
\end{itemize}
Example~\ref{exm:res} shows that the converse of this axiom is not possible. Besides, we can define another operator of \emph{individual} knowledge pooling:
\[
R^{a_1}_{G}\varphi:=[a_1\smalltriangleright a_2]\dots[a_{n-1}\smalltriangleright a_n]\varphi.
\]
Now we have the following relationship between individual knowledge, ``everybody knows''~\cite{van2007dynamic}, individual sharing, collective sharing, and information pooling. 
\begin{proposition} \label{prop:RG}
 These properties are sound:
\begin{enumerate}
    \item $K_{a_1}\varphi \to R^{a_1}_GE_G\varphi$ where $\varphi \in Prop^+$ \label{prop:rg1}
    \item $R^{a_1}_GE_G\varphi \to R_GE_G\varphi$ where $\varphi \in Prop^+$  \label{prop:rg2}
    \item $E_G\varphi \to R_GE_G\varphi$ where $\varphi \in Prop^+$  \label{prop:rg3}
    \item $R_GE_G\varphi \to R^i_GE_G\varphi$ where $\varphi \in Prop^+$  \label{prop:rg4}
\end{enumerate}
\end{proposition}

\begin{example} \label{exm:res}
The model $M$ in Figure~\ref{fig:ks} is a counter-example of $R^i_GE_G\phi \to R_GE_G\phi$. In this model, the information resolution pools all consistent but unknown information gathering from agent $a$ and $b$ to have $R_a\cap R_b = \{(s_0,s_0)\}$. Therefore, 
\begin{itemize}
    \item $M, s_0 \models R^i_{\{a,b\}}E_{\{a,b\}}(p\wedge q \wedge r)$.
\end{itemize}
However, as both agents possess the same individual knowledge -- each knows $p$ but nothing else -- this results in: 
\begin{itemize}
    \item $M, s_0 \models R_{\{a,b\}}E_{\{a,b\}}p$, 
    \item $M, s_0 \not\models R_{\{a,b\}}E_{\{a,b\}}q$, and
    \item $M, s_0 \not\models R_{\{a,b\}}E_{\{a,b\}}r$.
\end{itemize}
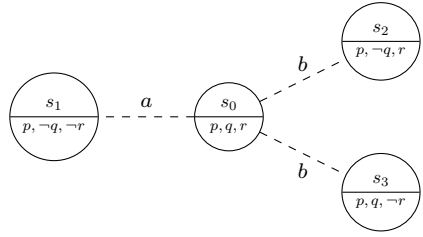
\begin{figure}
    \centering
    \input{TabFig/fig_ks}
    \caption{From information pooling to knowledge sharing.}
    \label{fig:ks}
\end{figure}
\end{example}

This Example~\ref{exm:res} demonstrates that knowledge pooling results in a distinct form of collective knowledge, diverging from the distributed knowledge generated through information resolution. The relationship among individual knowledge, the knowledge achieved by knowledge pooling, and the knowledge obtained by information pooling can be outlined as follows: For $\phi \in Prop^+$, 
\[
E_G\varphi \to K_a\phi \to R^a_GE_G\varphi \to R_GE_G\phi \to R^i_GE_G\phi \leftrightarrow D_GR^i_G\phi.
\]

An important aspect of knowledge pooling is that, despite the various directions in which it can occur, there always exists a resolution that obtains all knowledge from the participating agents, as illustrated in Example~\ref{exm:direction} below. 

\begin{example}[Sensitive Information, Cont'd] \label{exm:direction}
Starting with the initial model presented in Figure~\ref{fig:RG}, knowledge sharing towards the customer $c$ can proceed in two directions: either server $a$ or server $b$ initiates sharing their knowledge. This leads to distinct updated models, as shown in Figure~\ref{fig:2ks}:
\begin{itemize}
    \item $M, s_0 \models [a\smalltriangleright c]K_c (p\to q)$;
    \item $M, s_0 \models [b\smalltriangleright c]K_c(q\to r)$.
\end{itemize}
   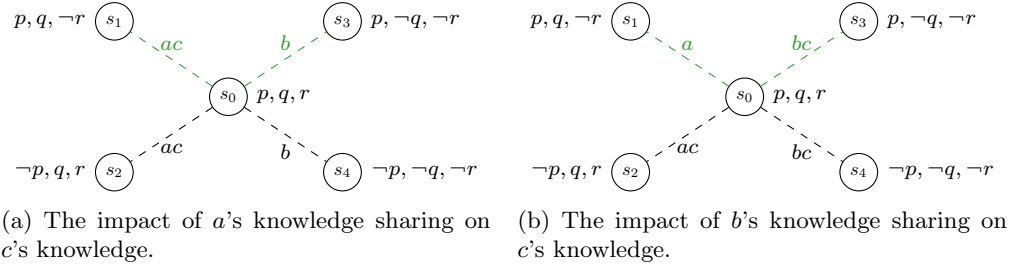
\begin{figure}
 \subcaptionbox{The impact of $a$'s knowledge sharing on $c$'s knowledge. }{\input{TabFig/fig_dire1.tex}}\quad
 \subcaptionbox{The impact of $b$'s knowledge sharing on $c$'s knowledge.}{\input{TabFig/fig_dire2.tex}}
    \caption{Two different directions of knowledge sharing}
    \label{fig:2ks}
\end{figure}

Within the framework of knowledge resolution, the concept of ``everybody knows'' knowledge~\cite{van2007dynamic} is achievable: 
\begin{itemize}
    \item $M, s_0 \models R^a_{\{a,b,c\}}E_{\{a,b,c\}}(p\to q)$;
    \item $M, s_0 \not\models R^a_{\{a,b,c\}}E_{\{a,b,c\}}(q \to r)$;
    \item $M, s_0 \models R^b_{\{a,b,c\}}E_{\{a,b,c\}}(q\to r)$;
    \item $M, s_0 \not\models R^b_{\{a,b,c\}}E_{\{a,b,c\}}(p\to q)$;
    \item $M, s_0 \models R_{\{a,b,c\}}(E_{\{a,b,c\}}(p\to q) \wedge E_{\{a,b,c\}}(q\to r) \wedge E_{\{a,b,c\}}(p\to r))$.
\end{itemize}
\end{example}

The concept of knowledge resolution highlights a crucial insight from knowledge pooling: its capacity to facilitate a consensus among a group~\footnote{If we incorporate the notion of common knowledge~\cite{van2007dynamic} into our framework, we can validate the following statement about our knowledge resolution:
\[
R_GE_G\varphi \leftrightarrow R_GC_G\varphi,
\]
where $C_G$ is the modality of common knowledge in group $G$.}. However, a knowledge resolution may not always be allowed, particularly in scenarios involving sensitive information or issues arising from epistemic implications. The next section will propose a formalization for \emph{permissible} knowledge pooling to address the gap between epistemic and ethical implications.

\section{Permissible Knowledge Pooling} \label{sec:perm}

\subsection{Permission to Know: Static Norms}

To capture ethical implications within our model, we incorporate the concept of \emph{ideal knowledge transition}. In standard deontic logic, the Kanger reduction typically employs a propositional constant to identify states within the model as permissible or impermissible~\cite{gabbay2013handbook}. To examine the interplay between epistemic and ethical implications, we introduce a propositional constant that specifies which pairs of states in the model, referred to as knowledge transitions, are permissible. Those knowledge transitions considered permissible are thus classified as ideal knowledge transitions.  

We expand our framework of knowledge pooling as described below. A structure $M = \langle W, \{R_a\}_{a\in \agt}, O, V \rangle$ is a deontic model when $M$ is a model in additional with a symmetric relation $O \subseteq W \times W$ with the property that $\emptyset \neq O \subseteq \bigcup_{a\in\agt}R_a$. To represent an ideal transition, we introduce the constant ${\sf O}$ and extend our language, indicating the presence of a \emph{permissible} way to know -- suggesting that it is \emph{okay} to gain knowledge in such a manner. This constant is interpreted at a point $w$ in a deontic model $M$ as follows. 
\begin{align*}
    M, w \models {\sf O} \textrm{ iff }  O[w]\neq \emptyset.
\end{align*}
This constant has this validity:
\begin{center}
    ${\sf O} \to \hat{K_a}{\sf O}$
\end{center}
for some agent $a$.~\footnote{By incorporating the concept of common knowledge, the validity expressed by $C_{\agt}\bigvee_{a \in \agt}O$ captures the characteristics of $O \subseteq \bigcup_{a\in\agt}R_a$ and $O \neq \emptyset$.}

\begin{example}[Sensitive Information, Cont'd]
Figure~\ref{fig:RG} illustrates the state pairs $(s_1,s_0)$ and $(s_3,s_0)$ along with their corresponding symmetric pairs, constitute the ideal relation $O$. These pairs are thus considered as permissible pathways of obtaining knowledge.  
\end{example}

We introduce a \emph{static} notion termed as weak permission, denoted by $P_a\varphi$, and define it as $K_a\varphi \wedge \hat{K_a}{\sf O}$. Thus, $P_a\varphi$ at a point $w$ in the deontic model $M$ is interpreted as:
\begin{align*}
    M, w \models P_a\varphi \textrm{ iff } R_a[w]\subseteq ||\varphi|| \textrm{ and } R_a[w] \cap O[w] \neq \emptyset
\end{align*}
This implies that when an agent $a$ is permitted to acquire knowledge, it means that there is at least one ideal way for her to access her knowledge. The formula $P_a\varphi$ is designed to captures the concept of having permission to know. It asserts that an agent $a$ is permitted to know $\varphi$ if and only if, $\varphi$ is part of agent $a$'s knowledge base, and there is at least one permissible way of knowing it. This concept of ``permission to know'' can be expressed using a static language. 

\begin{example}[Sensitive Information, Cont'd] \label{exm:p1}
After server $a$ conveys her knowledge to customer $c$, it becomes permissible for the agent $c$ to know $p \to q$. Likewise, when server $b$ shares his knowledge with customer $c$, it becomes permissible for $c$ to know $q \to r$. These permissions are true at point $s_0$ in the models illustrated in Figure~\ref{fig:RG}:
\begin{itemize}
    \item $M|_{a\smalltriangleright c}, s_0 \models P_c(p\to q)$;
    \item $M|_{b\smalltriangleright c}, s_0 \models P_c(q \to r)$.
\end{itemize}
However, in the model in Figure~(\ref{fig:2ks}a), the green and unique $c$-link will be deleted after the $(b\smalltriangleright c)$-update, so it is not permissible for $c$ to know $p\to r$.  
\begin{itemize}
    \item For instance, $(M|_{a\smalltriangleright c})|_{b\smalltriangleright c}, s_0 \not\models P_c(p\to r)$.
\end{itemize} 
\end{example}

The static concept of permission to know displays characteristics derived from the principles of free choice permission~\cite{hansson2013varieties} and individual knowledge:  
\begin{proposition}
Here are some validities:
\begin{enumerate}
    \item {\sf RFC}: $P_a\varphi \wedge P_a\psi \to P_a(\varphi \vee \psi)$
    \item {\sf MC}: $P_a(\varphi \wedge \psi) \leftrightarrow P_a\varphi \wedge P_a\psi$
    \item {\sf K}: $P_a(\varphi \to \psi) \to (P_a\varphi \to P_a\psi)$
    \item {\sf D}: $\neg P_a\bot$
    \item {\sf T}: $P_a\varphi \to \varphi$ 
    \item {\sf 4}: $P_a\varphi \to P_aP_a\varphi$ 
    \item {\sf RE}: From $ \varphi \to \psi$ it infers $ P_a\varphi \to P_a\psi$
    \item {\sf NEC}: From $ \varphi$ it infers $ P_a\varphi$
\end{enumerate}
\end{proposition}

However, this concept avoids the infamous free choice permission paradox~\cite{hansson2013varieties} and the {\sf 5} axiom: 
\begin{example}
Here are some invalidities for all $a \in \agt$:
\begin{enumerate}
    \item {\sf FCP$_1$}: $P_a(\varphi \vee \psi) \to P_a\varphi \wedge P_a\psi$
    \item {\sf FCP$_2$}: $P_a\varphi \to P_a(\varphi \wedge \psi)$
    \item {\sf 5}: $\hat{P_a}\varphi \to P_a\hat{P_a}\varphi$ 
\end{enumerate}
\end{example}

Significantly, the permission to access knowledge represents a novel bridge linking individual knowledge with distributed knowledge, as described below:
\begin{align*}
      K_a\varphi\to P_a\varphi \to D_G(\varphi \wedge {\sf O}).  
\end{align*}

The dual concept of ``permission to know'' -- ``ought to know'' can be defined in the usual manner: $O_a\varphi: = \hat{P_a}\varphi$. Thus, ``ought to know'' $O_a\varphi$ corresponds to a conditional $ (K_a\neg\varphi \to K_a\neg{\sf O})$. This interpretation of ``ought to know'' avoids the infamous \r{A}qvist's paradox~\cite{aaqvist1967good}.

\subsection{Permission to Knowledge Pooling: Dynamic Norms}

Now we define a new type of permission -- the permission for one to communicate their knowledge towards another. This introduces a novel \emph{dynamic} dimension to the concept of permission, particularly in terms of communication, that is represented as $P(a \smalltriangleright b)$. It denotes that agent $a$ is permitted to share their knowledge with that of agent $b$. More precisely, 
\[
P(a\smalltriangleright b)=_{df}[a\smalltriangleright b]\hat{K_b}{\sf O}
\]
Agent $a$ is allowed to transfer knowledge to agent $b$, if and only if 
the subsequent communication from $a$ to $b$ results in a permissible way for receiver $b$ to now. 


\begin{example}[Sensitive Information, Cont'd]
Within the use of the formulas like $P(a\smalltriangleright b)$, we can further explore the case of sensitive information in Example~\ref{exm:p1}. In the model $M|_{a\smalltriangleright c}$, server $b$ still knows $q \to r$. When $b$ shares this knowledge with customer $c$, $c$ knows it as well. However, there is no legitimate means for her to having obtained this knowledge, as shown in Example~\ref{exm:p1}: 
\begin{itemize}
    \item $M|_{a\smalltriangleright c}, s_0 
    \not\models [b\smalltriangleright c]\hat{K_c}{\sf O}$.
\end{itemize}
From these observations, we deduce that the transfer of knowledge from $b$ to $c$ is impermissible. 
\begin{itemize}
    \item $M|_{a\smalltriangleright c}, s_0 \not\models P(b \smalltriangleright c)$.
\end{itemize}
\end{example}

Knowledge pooling shares significant ties with the concept of ``permissions to share knowledge'', particularly within multi-agent interaction scenarios.
\begin{proposition} \label{prop:perm}
    Here are two validities:
\begin{enumerate}
  \item $[a\smalltriangleright b]P(b\smalltriangleright c) \to P(a\smalltriangleright c)$ \label{prop:o-tran}
    \item \label{prop:Keq} If $\vdash K_a\varphi \leftrightarrow K_b \varphi$ then $P(a\smalltriangleright c) \leftrightarrow P(b \smalltriangleright c)$
\end{enumerate}
\end{proposition}

Proposition~\ref{prop:perm}.\ref{prop:o-tran} demonstrates that, if it is permissible to share knowledge from $b$ to $c$ after $a$'s sharing towards $b$, a knowledge sharing from $a$ directly to $c$ is also permissible. Proposition~\ref{prop:perm}.\ref{prop:Keq} addresses the condition for replacing participants in permissible pooling: if the originating knowledge of the senders is identical, then the permissibility status of their knowledge pooling is the same. Therefore, Proposition~\ref{prop:perm}.\ref{prop:Keq}  explains one scenario in which epistemic implications can affect ethical considerations. Example~\ref{exm:norep} below shows that substituting recipients with identical knowledge does not guarantee the preservation of the original permissibility status.    
\begin{example} \label{exm:norep}
    Here are some invalidities:
\begin{enumerate}
    \item If $K_b\varphi \leftrightarrow K_c\varphi$ then $P(a\smalltriangleright b) \leftrightarrow P(a \smalltriangleright c)$.
\end{enumerate}
\end{example}

\section{Expressivity, Axiomatization, and Completeness} \label{sec:tech}

This section presents several technical results, including the expressivity, soundness, and completenss of various axiomatizations of knowledge sharing. 

\subsection{Expressivity} \label{sec:express}

Now we list different langauges of knowledge sharing as follows:
\begin{align*}
    \mathcal{L}_{IK}::\,&\varphi:= p \mid \varphi\wedge\varphi \mid K_a\varphi\\
    \mathcal{L}_{AK}::\,&\varphi:= p  \mid \varphi\wedge\varphi \mid K_a\varphi \mid K^a_b\varphi\\
    \mathcal{L}_{DK}::\,&\varphi:= p  \mid \varphi\wedge\varphi \mid K_a\varphi \mid K^a_b\varphi \mid D_G\varphi \\
    \mathcal{L}_{IKS}::\,&\varphi:= p \mid \varphi\wedge\varphi \mid K_a\varphi \mid [a\smalltriangleright b]\varphi\\
    \mathcal{L}_{AKS}::\,&\varphi:= p  \mid \varphi\wedge\varphi \mid K_a\varphi \mid K^a_b\varphi \mid [a\smalltriangleright b]\varphi\\
    \mathcal{L}_{DKS}::\,&\varphi:= p  \mid \varphi\wedge\varphi \mid K_a\varphi \mid K^a_b\varphi \mid D_G\varphi \mid [a\smalltriangleright b]\varphi\\
    \mathcal{L}_{DKR}::\,&\varphi:= p  \mid \varphi\wedge\varphi \mid K_a\varphi \mid K^a_b\varphi \mid D_G\varphi \mid R^i_G\varphi \mid [a\smalltriangleright b]\varphi
\end{align*}
where $p$ is an element of the (countable) set $Prop$ of atomic propositions, $a,b \in \mathcal{I}$ are elements of the (finite) set $\mathcal{I}$ of agents, and $G \subseteq \mathcal{I}$ is a subset of agents. 

\begin{theorem}
  The diagram in Figure~\ref{fig:express} illustrates the expressive power, indicating that any language pointed to by an arrow is strictly more expressive than the language from which the arrow originates.
\begin{figure}
    \centering
    \input{TabFig/fig_express}
    \caption{Expressivity}
    \label{fig:express}
\end{figure}
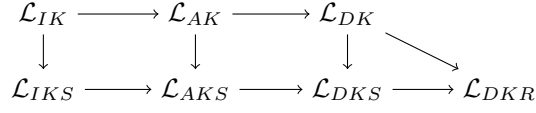
\end{theorem}

\subsection{System {\bf AK} of Agent-dependent Knowledge} \label{sec:ak}

The logical system {\bf AK} of agent-dependent knowledge is presented in Figure~\ref{fig:ak}. It includes two interaction axioms {\sf Int} and {\sf Cl} to connect between individual and agent-dependent knowledge.  

\begin{figure}
    \centering
    \fbox{
\begin{tabular}{ll}
    ({\sf S5}) & The S5 proof system for individual knowledge~\cite{van2007dynamic} \\
    ({\sf AK}) & S5 axioms for agent-dependent knowledge\\
    ({\sf Int}) & $K_a\varphi \to K^b_a\varphi$\\
    ({\sf Cl}) & $K^b_a\varphi\to\bigvee_{\psi \in \mathcal{L}}(K_a\psi \wedge K_b(\psi\to\varphi))$\\
    ({\sf NEC$_A$}) & From $\varphi$ inferring $K^b_a\varphi$
\end{tabular}}
    \caption{Axiomatization {\bf AK}}
    \label{fig:ak}
\end{figure}

\begin{theorem}
    The system {\bf AK} is sound and (strongly) complete.
\end{theorem}

\begin{proposition}
    The axiom {\sf Cl} is valid. 
\end{proposition}
\begin{proof}
    Let $M, w \models K^b_a\varphi$. This is $(R_a \cap \equiv_{b:w})[w]\subseteq||\varphi||$. Now we need to find a $\psi \in \mathcal{L}_{AK}$ to satisfy $R_a[w]\subseteq ||\psi \to\varphi||$ and $R_b[w] \subseteq ||\psi||$. Notice that $R_b[w]\subseteq\equiv_{b:w}[w]$ by Lemma~\ref{lem:update}.\ref{lem:u4} and $R_b[w] \neq \emptyset$. So we are able to find $\psi \in \mathcal{L}_{AK}$ such that $R_b[w] \subseteq ||\psi||\subseteq \equiv_{b:w}[w]$. Now we need to show $R_a[w]\subseteq ||\psi\to\varphi||$. If not, there is a $u\in R_a[w]$ such that $u \in ||\psi \wedge \neg\varphi||$. However, from $u\in ||\psi|| \subseteq \equiv_{b:w}[w]$ and $u \in R_a[w]$, it infers $u \in ||\varphi||$. This leads to a contradiction.           
\end{proof}

\begin{definition}[Maximally Consistent Sets]
    Given the system {\bf AK}, for any $\Sigma \subseteq \mathcal{L}_{AK}$:
\begin{itemize}
    \item $\Sigma$ is {\bf AK}-consistent iff each finite $X \subseteq \Sigma$ such that $\not\vdash_{\bf AK} \neg \bigwedge X$; otherwise $\Sigma$ is {\bf AK}-inconsistent.
    \item $\Sigma$ is a maximally {\bf AK}-consistent subset iff it is {\bf AK}-consistent and if $\Sigma \subset \Sigma' \subseteq \mathcal{L}_{AK}$ then $\Sigma'$ is {\bf AK}-inconsistent. 
\end{itemize}
We define $MCS$ be the set of all maximally {\bf AK}-consistent subsets. 
\end{definition}

\begin{lemma}
    Let $\Sigma$ be a member of $MCS$. For any $\varphi, \psi \in \mathcal{L}_{AK}$: 
\begin{itemize}
    \item $\varphi \not\in \Sigma \Leftrightarrow \neg\varphi \in \Sigma$.
    \item $\varphi \wedge \psi \in \Sigma \Leftrightarrow \varphi, \psi \in \Sigma$.
    \item $\Sigma$ is closed under the logic {\bf AK}, such that 
    \begin{center}
        $\varphi \in \Sigma$ and $\vdash_{\bf AK}\varphi \to \psi$ imply $\psi \in \Sigma$.
    \end{center}
\end{itemize}
\end{lemma}

\begin{definition}[Canonical Model]
    Given the system {\bf AK}, we construct the canonical model $M^C = \langle W^C, \{R^C_a\}_{a\in \agt}, V^C\rangle$ as follows:
\begin{itemize}
    \item $W^C = MCS$ is all maximally {\bf AK}-consistent subsets. 
    \item $R^C_a = \{(\Delta, \Theta) \subseteq W^C \times W^C\mid K_a\varphi \in \Delta \Rightarrow \varphi \in \Theta\}$
    \item $V^C(p) = \{\Delta \in W^C \mid p \in \Delta\}$
\end{itemize}
\end{definition}
\noindent
Hence, $\equiv_{b:\Gamma}^C$ is defined as $\{(\Delta, \Theta) \subseteq W^C \times W^C\mid \forall \varphi \in K_b\Gamma(\varphi \in \Delta \Leftrightarrow \varphi \in \Theta)\}$. Concerns about the possibility of an empty intersection~\cite{castaneda2023communication} between $R^C_a \cap \equiv_{b:\Gamma}^C$ do not arise in our canonical model. 

\begin{lemma}[Truth Lemma]
Let $M^C = \langle W^C, \{R^C_a\}_{a\in \agt}, V^C\rangle$ be the canonical model with respect to logic {\bf AK}. For any $\varphi \in \mathcal{L}_{AK}$ and $\Delta \in W^C$:
\[
M^C, \Delta \models \varphi \text{ iff } \varphi \in \Delta.
\]
\end{lemma}
\begin{proof}
    We only focus on the case of agent-dependent knowledge $K^b_a\varphi$. The other cases are standard.
\begin{itemize}
    \item Suppose $K^b_a\varphi \in \Delta$. We want to show $(R^C_a \cap \equiv^C_{b:\Delta})[\Delta]\subseteq ||\varphi||_{M^C}$. Let $\Theta \in (R_a \cap \equiv_{b:\Delta})[\Delta]$. From $K^b_a\varphi \in \Delta$ and {\sf Cl} axiom they infer $K_a\psi \in \Delta$ and $K_b(\psi \to \varphi)\in \Delta$ for some $\psi\in\mathcal{L}_{AK}$. From $(\Delta,\Theta)\in R^C_a$ it then infers $\psi \in \Theta$. From $(\Delta,\Theta)\in \equiv_{b:\Delta}$ it infers $\psi\to\varphi\in\Theta$. So $\varphi\in \Theta$. By inductive hypothesis, it is $\Theta \in ||\varphi||_{M^C}$. 
    \item Suppose $\neg K^b_a\varphi \in \Delta$. We need a $\Theta\in MCS$ such that $\Theta \in (R^C_a \cap \equiv^C_{b:\Delta})[\Delta]$ but $\Theta \not\in ||\varphi||_{M^C}$. We construct such a $\Theta$ in the following way. We enumerate the formulas in the countable set $\mathcal{L}_{AK}$ as $\varphi_1, \dots, \varphi_n, \dots$ and construct $\Gamma_i$ for $i \in \mathbb{N}$ as follows:
    \begin{align*}
        \Gamma_0 & = \{\neg\varphi\} \cup \Delta_0,\\
        \Gamma_{i+1} & = 
        \begin{cases}
      \Gamma_i \cup \{\varphi_i \mid K_a^b\varphi_i \in \Delta\}& \text{if this set is {\bf AK}-consistent; }\\
      \Gamma_i & \text{else. }
    \end{cases}
    \end{align*}
    where $\Delta_0 = \{\psi \in \mathcal{L}_{AK} \mid K_b\psi\in \Delta \}$.
    \begin{enumerate}
        \item We need to show $\Gamma_i$ where $i \in \mathbb{N}$ is consistent. 
        \begin{enumerate}
        \item $\Delta_0$ must be consistent. We want to show $\Gamma_0$ is consistent. If not, then it must be some $\psi_1, \dots, \psi_n \in \Delta_0$ such that $\vdash \bigwedge_{1 \leq i \leq n}\psi_i \to \varphi$. By the {\sf RK} rule of $K^b_a$ modality, it infers $\vdash \bigwedge_{1 \leq i \leq n}K^b_a\psi_i \to K^b_a\varphi$. Due to $K^b_a\psi_i \in \Delta$ for any $1 \leq i \leq n$ and $\Delta \in MCS$, it leads to $K^b_a\varphi \in \Delta$. This leads to a contradiction.    
        \item Every $\Gamma_{i+1} = \Gamma_i \cup \{\neg\varphi\}$ should be consistent. Otherwise, there are $\psi_1, \dots, \psi_n \in \Gamma_i$ such that $\vdash \neg\varphi \to \neg\bigwedge_{1 \leq i \leq n}\psi_i$. That is $\vdash \bigwedge_{1 \leq i \leq n}\psi_i \to \varphi$. By the rule {\sf RK} of $K^b_a$ modality, it infers $\vdash \bigwedge_{1 \leq i \leq n}K_a^b\psi_i \to K_a^b\varphi$. Notice $\neg K^b_a\varphi \in \Delta$. By the closure of $\vdash$ in $MCS$, it infers $\neg \bigwedge_{1 \leq i \leq n}K_a^b\psi_i \in \Delta$. However, $K_a^b\psi_i \in \Delta$ for each $1 \leq i \leq n$. It leads to a contradiction that $\Delta$ is consistent.  
        \end{enumerate}
        \item We define $\Gamma = \bigcup_{i\in \mathbb{B}}\Gamma_i$. We need to show $\Gamma \in MCS$. To verify that $\Gamma$ is consistent, the proof is similar to the previous case. The next is to show $\Gamma$ is maximal. If not, then there is a $\psi$ such that $\Gamma \cup \{\psi\} \supset \Gamma$ is still consistent. In such a case, $\Gamma \cup \{\psi\} \subseteq \Gamma$, which leads to a contradiction.  
        \item We want to show $\Gamma \in (R^C_a \cap \equiv^C_{b:\Delta})[\Delta]$, which is
        \begin{enumerate}
            \item $\forall \varphi \in \mathcal{L}_{AK}\,(K_a\varphi \in \Delta \Rightarrow \varphi \in \Gamma)$, and  \label{complete:know}
            \item $\forall \varphi \in K_b\Delta \,(\varphi \in \Delta \Leftrightarrow \varphi \in \Gamma)$ \label{complete:eqv}
        \end{enumerate}
        From the {\sf Int} axiom, given any $K_a\varphi \in \Delta$ it infers $K^b_a\varphi \in \Delta$. By the construction of $\Gamma$, it gives $\varphi \in \Gamma$. So the condition~(\ref{complete:know}) is verified. The condition~(\ref{complete:eqv}) can be verified by the construction of $\Gamma$. Given any $K_b\varphi \in \Delta$, it is the case that $\varphi \in K_b\Delta$. By the {\sf T} axiom of $K_b$ modality it is always $\varphi \in \Delta$, while by the construction of $\Gamma$ it is the case that $\varphi \in \Gamma$. So the condition~(\ref{complete:eqv}) is always satisfied.    
    \end{enumerate}
\end{itemize}
\end{proof}

\subsection{System {\bf IKS} of Knowledge Sharing} \label{sec:iks}

The dynamic logic {\bf IKS} of knowledge sharing is presented in Figure~\ref{fig:iks}. Notably, this system adopts a non-reductionist approach~\cite{wang2013axiomatizations} to model knowledge sharing.  

\begin{figure}
    \centering
    \fbox{
\begin{tabular}{ll}
    ({\sf S5}) & The proof system of individual knowledge\\
    ({\sf Inv}) & $(p \to [a\smalltriangleright b]p) \wedge (\neg p \to [a\smalltriangleright b]\neg p)$\\
    ({\sf Int$^+$}) & $[a\smalltriangleright b]K_b\varphi \to \bigvee_{\psi\in\mathcal{L}}(K_b[a\smalltriangleright b](\psi \to \varphi) \wedge K_a[a\smalltriangleright b]\psi )$\\
    ({\sf Int$_+$})& $K_b[a\smalltriangleright b]\varphi \to [a\smalltriangleright b]K_b\varphi$\\
    ({\sf Int$^-$}) & $[a\smalltriangleright b]K_c\varphi \leftrightarrow K_c[a\smalltriangleright b]\varphi$ where $b \neq c$\\
    ({\sf Rev})& $\neg[a\smalltriangleright b]\varphi \to [a\smalltriangleright b]\neg\varphi$\\
    ({\sf D})& $[a\smalltriangleright b]\neg\varphi \to \neg[a\smalltriangleright b]\varphi$ \\
    ({\sf K$_{\smalltriangleright}$}) & $[a\smalltriangleright b](\varphi \to \psi) \to ([a \smalltriangleright b]\varphi \to [a\smalltriangleright b]\psi)$ \\
    ({\sf Rep})& $[a\smalltriangleright b]\varphi \leftrightarrow [a\smalltriangleright b][a\smalltriangleright b]\varphi$\\
    ({\sf N$_s$}) & from $\varphi$ infer $[a\smalltriangleright b]\varphi$\\
    ({\sf Inc$_{\smalltriangleright}$}) & from $\varphi \to [a\smalltriangleright b]\psi$ infer $K_a\varphi \to [a\smalltriangleright b]K_b\psi$
\end{tabular}}
    \caption{Axiomatization {\bf IKS} with non-reduction axioms}
    \label{fig:iks}
\end{figure}

\begin{theorem}
    The system {\bf IKS} is sound and (strongly) complete. 
\end{theorem}

\begin{proposition}
    The axioms and rules in {\bf IKS} are valid. 
\end{proposition}
\begin{proof}
Now we prove the validities of all axioms and rules for the $[a\smalltriangleright b]$ modality. 
\begin{itemize}
    \item The validity of the {\sf Inv} axiom can be verified by the definition of updated model. 
    \item Suppose $M, w\models [a\smalltriangleright b]K_b\varphi$. This is equal to $R_b^{a\smalltriangleright b}[w] \subseteq ||\varphi||^*$. Now we need to find a $\psi \in \mathcal{L}_{IKS}$ such that $R_b[w]\subseteq ||\psi \to\varphi||^*$ and $R_a[w]\subseteq || \psi||^*$. As $R_a[w] \subseteq \equiv_{a:w}[w]$ and $R_a[w] \neq \emptyset$, we are able to find out such a $\psi \neq \bot$ such that $R_a[w] \subseteq  ||\psi||^* \subseteq \equiv_{a:w}[w]$. Now we need to show $R_b[w]\subseteq ||\psi\to\varphi||^*$. If not, there is a $u\in R^b[w]$ such that $u \in ||\psi \wedge \neg\varphi||^*$. From $u \in ||\psi||^* \subseteq \equiv_{a:w}[w]$ and $u \in R^b[w]$, it follows that $u \in R^{a\smalltriangleright b}_b \subseteq ||\varphi||^*$. It then leads to a contradiction. So the {\sf Int$^+$} axiom is valid.  

\item This is straightforward by considering the definition of update.  

\item Due to the definition of update, {\sf Int$^-$} is valid.

\item From $M, w \not\models [a\smalltriangleright b]\varphi$ it infers $M|_{a\smalltriangleright b}, w \not\models \varphi$. This is $M|_{a\smalltriangleright b}, w \models\neg\varphi$. So $M, w \models [a\smalltriangleright b]\neg\varphi$.

\item Suppose $M, w \models [a\smalltriangleright b]\bot$. So $M|_{a\smalltriangleright b}, w \models \bot$, which is not possible. 

\item Suppose $M|_{a\smalltriangleright b}, w \models \varphi \to \psi$ and $M|_{a\smalltriangleright b}, w \models \varphi$. So $M|_{a\smalltriangleright b}, w \models \psi$. So the {\sf K$_{\smalltriangleright}$} axiom holds. 

\item Suppose $M|_{a\smalltriangleright b}, w \models \varphi$. Then, update the model $M|_{a\smalltriangleright b}$ by $a\smalltriangleright b$ does not change any truth assignments or binary relations in this model. So $(M|_{a\smalltriangleright b})|_{a\smalltriangleright b}, w \models \varphi$.  

\item Because $\varphi$ is valid. So, as in the standard proof~\cite{van2007dynamic}, $\varphi$ is also valid in the updated models. 

\item Suppose $||\varphi ||\subseteq ||\psi||^{a\smalltriangleright b}$. Let $R_a[w]\subseteq ||\varphi||$ and $u\in R^{\smalltriangleright}_{b:c}[w]$. From $u\in R^{\smalltriangleright}_{b:c}[w] \subseteq \equiv_{a:w}$ it infers that $u \in ||\psi|| \Leftrightarrow w \in ||\psi||$ for any $\psi \in K_a w$. From $R_a[w]\subseteq ||\varphi||$ it leads to $u \in ||\varphi||$. From the assumption $||\varphi ||\subseteq ||\psi||^{a\smalltriangleright b}$ it follows $u \in ||\psi||^{a\smalltriangleright b}$. Now we have $R^{\smalltriangleright}_{b:c}[w] \subseteq ||\psi||^{a\smalltriangleright b}$. So the {\sf Inc$_{\smalltriangleright}$} rule is valid. 

\end{itemize}
\end{proof}

\begin{proposition}
    These are valid in {\bf IKS}:
\begin{itemize}
\item {\sf C}: $[a\smalltriangleright b]\varphi \wedge [a\smalltriangleright b]\psi \to [a\smalltriangleright b](\varphi \wedge \psi)$
\item {\sf RK}: From $\vdash \varphi \wedge \varphi '\to \psi$, it infers $\vdash [a\smalltriangleright b]\varphi \wedge [a\smalltriangleright b]\varphi' \to [a\smalltriangleright b]\psi$.

\item {\sf RM$_{\smalltriangleright}$}: From $\varphi_1,\dots, \varphi_n \to [a\smalltriangleright b]\psi$ inferring $K_c\varphi_1,\dots, K_c\varphi_n \to [a\smalltriangleright b]K_c\psi$.
\end{itemize}
\end{proposition}
\begin{proof}
\begin{itemize}
    \item The axioms and rules {\sf NEC$_{K_c}$, Int$^-$}, and {\sf Int$_+$} lead to the validity of the {\sf RM$_{\smalltriangleright}$} rule. 
\end{itemize}
\end{proof}

\begin{proposition}
 These are derivable in {\bf IKS}.
\begin{enumerate}
 \item ({\sf Boolean}) $\varphi \leftrightarrow [a\smalltriangleright b]\varphi$ where $\varphi \in Prop^+$
 \item ({\sf Remain}) $K_c\varphi \to [a\smalltriangleright b]K_c\varphi$ where $\varphi \in Prop^+$
 \item ({\sf Sharing}) $K_a\varphi \to [a\smalltriangleright b]K_b\varphi$ where $\varphi \in Prop^+$
 \item ({\sf Step}) $[a\smalltriangleright b]K_b\varphi \to [a\smalltriangleright b][b\smalltriangleright c]K_c\varphi$ where $\varphi \in Prop^+$
\end{enumerate}   
\end{proposition}
\begin{proof}
\begin{enumerate}
     \item {\sf Boolean} can be derived by {\sf Inv, D, K$_{\smalltriangleright}$}, and {\sf N$_s$}.
    \item {\sf Remain} can be verified by {\sf Boolean}, and {\sf RM$_{\smalltriangleright}$}.  
    \item {\sf Sharing} can be verified by {\sf Boolean} and {\sf Inc$_{\smalltriangleright}$}.
    \item {\sf Step} can be verified by {\sf Sharing, K$_{\smalltriangleright}$, N$_s$}, and {\sf MP}.
   \end{enumerate}
\end{proof}

We adopt the \emph{detour} canonical method~\cite{wang2013axiomatizations} to prove the completenss of this dynamic logic {\bf IKS}.

\begin{definition}[Extended Model] An extended model $M$ for {\bf IKS} is a structure $\langle W, \{R_a\}_{a \in \agt}, \{R_{a\smalltriangleright b}\}_{a,b\in\agt},V\rangle$ where:
\begin{itemize}
    \item $(\langle W, \{R_a\}_{a \in \agt},V\rangle)$ is a model for {\bf IKS};
    \item For any $a,b\in \agt$, $R_{a\smalltriangleright b}$ is a relation over $W$.
\end{itemize}
\end{definition}
\noindent
We call $(\langle W, \{R_a\}_{a \in \agt},V\rangle)$ the core of the extended model $M$ and denote it as $M^-$. 

The truth conditions are defined as follows:
\begin{align*}
    M,w\models^{\smalltriangleright} p &\textrm{ iff } w \in V(p)\\
    M,w \models^{\smalltriangleright} \neg\varphi &\textrm{ iff } M, w\not\models^{\smalltriangleright}\varphi\\
    M, w \models^{\smalltriangleright} \varphi \wedge \psi &\textrm{ iff } M, w \models^{\smalltriangleright}\varphi \textrm{ and } M, w \models^{\smalltriangleright} \psi\\
    M, w \models^{\smalltriangleright} K_a\varphi &\textrm{ iff } \forall u \in R_a[w]: M,u\models^{\smalltriangleright} \varphi\\
    M, w \models^{\smalltriangleright} [a\smalltriangleright b]\varphi &\textrm{ iff } \forall u \in R_{a\smalltriangleright b}[w]: M, u \models^{\smalltriangleright} \varphi.
\end{align*}

\begin{definition}[Normal extended model] An extended model $M = \langle W, \{R_a\}_{a \in \agt}, \{R_{a\smalltriangleright b}\}_{a,b\in\agt},V\rangle$ for {\bf IKS} is called \emph{normal}, if and only if for any $w, u \in M$ and any $a, b, c \in \agt$ such that $a \neq c$:
\begin{itemize}
    \item (U-Functionality) For any $w \in W$, $w$ has a unique $(a\smalltriangleright b)$-successor, that is $|R_{a\smalltriangleright b}[w]| = 1$. 
    \item (U-Invariance) If $wR_{a\smalltriangleright b} u$ then $\forall p \in Prop$: $w \in V(p) \Leftrightarrow u\in V(p)$.
    \item (U-Zig$^-$) If $wR_au$, $w R_{b\smalltriangleright c} w'$, and $uR_{b\smalltriangleright c}u'$, then $w'R_au'$. 
    \item (U-Zig$^+$) If $wR^b_cu$, $w R_{b\smalltriangleright c} w'$, and $uR_{b\smalltriangleright c}u'$, then $w'R_cu'$.
    \item (U-Zag$^-$) If $w'R_au'$ and $wR_{b\smalltriangleright c}w'$, then $\exists u$ such that $wR_au$ and $uR_{b\smalltriangleright c} u'$.
    \item (U-Zag$^+$) If $w'R_cu'$ and $wR_{b\smalltriangleright c}w'$, then $\exists u$ such that $wR^b_cu$ and $uR_{b\smalltriangleright c} u'$.
\end{itemize}
\end{definition}

\begin{definition}[Bisimulation] A binary relation $Z$ is called a $({a\smalltriangleright b})$-bisimulation between two pointed model $M,w$ and $N, u$, denoted as $M, w {\overset{a\smalltriangleright b}{\leftrightarroweq}} N, u$, when $wZu$ the following conditions of $w'Zu'$ hold:
\begin{itemize}
    \item (Invariance) $p \in V^M(w)$ iff $p \in V^N(u)$
    \item (Zig$^-$) If $wR_cu$ for some $u\in M$ then there is a $u' \in N$ such that $w'R_cu'$ and $uZu'$, where $c \neq b$.
    \item (Zig$^+$) If $wR^a_bu$ for some $u\in M$ then there is a $u' \in N$ such that $w'R_bu'$ and $uZu'$.
    \item (Zag$^-$) If $w'R_cu'$ for some $u'\in N$ then there is a $u \in M$ such that $wR_cu$ and $uZu'$, where $c \neq b$.
    \item (Zag$^+$) If $w'R_bu'$ for some $u'\in N$ then there is a $u \in M$ such that $wR^a_bu$ and $uZu'$.
\end{itemize}
\end{definition}

\begin{lemma} \label{lem:inv}
    Given a $\smalltriangleright$-normal extended model $M$: When $wR_{a\smalltriangleright b}
w'$ in $M$, 
    \[
    M^-|_{a\smalltriangleright b}, w \overset{a\smalltriangleright b}{\leftrightarroweq} M^-, w'.
    \]
\end{lemma}
\begin{proof}
    We need to show that $\overset{a\smalltriangleright b}{\leftrightarroweq}$ is a bisimulation. 
    \begin{itemize}
        \item (Invariance) That $p \in V^{M^-|_{a\smalltriangleright b}}(w) \Leftrightarrow p \in V^{M^-}(w')$ is ensured by U-Invariance. 
        \item (Zig$^-$) Let $wR_cu$ where $c\neq b$. By U-Functionality, there is a $u'$ such that $uR_{a\smalltriangleright b}u'$. By U-Zig$^-$, $w'R_cu'$. From $uR_{a\smalltriangleright b}u'$ and inductive hypothesis, it follows $M^-|_{a\smalltriangleright b}, u \overset{a\smalltriangleright b}{\leftrightarroweq} M^-, u'$.
        \item (Zig$^+$) Let $wR^a_bu$. By U-Functionality, there is a $u'$ such that $uR_{a\smalltriangleright b}u'$. By U-Zig$^+$, it infers $w'R_cu'$. From $uR_{a\smalltriangleright b}u'$ and inductive hypothesis, it follows $M^-|_{a\smalltriangleright b}, u \overset{a\smalltriangleright b}{\leftrightarroweq} M^-, u'$. 
        \item (Zag$^-$) Let $w'R_cu'$ where $c \neq b$. By U-Zag$^-$, there is a $u \in M^-$ such that $wR_cu$ and $uR_{a\smalltriangleright b}u'$. From $uR_{a\smalltriangleright b}u'$ and inductive hypothesis, it follows $M^-|_{a\smalltriangleright b}, u \overset{a\smalltriangleright b}{\leftrightarroweq} M^-, u'$. 
        \item (Zag$^+$) Let $w'R_bu'$. From U-Zag$^+$, there is a $u$ such that $wR^a_bu$ and $uR_{a\smalltriangleright b}u'$. From $uR_{a\smalltriangleright b}u'$ and inductive hypothesis, it follows $M^-|_{a\smalltriangleright b}, u \overset{a\smalltriangleright b}{\leftrightarroweq} M^-, u'$. 
    \end{itemize} 
\end{proof}

\begin{lemma} 
\label{lem:eq}
    When $M^-|_{a\smalltriangleright b}, w \overset{a\smalltriangleright b}{\leftrightarroweq} M^-, w'$, it is the case that 
    \[
    M^-|_{a\smalltriangleright b}, w \models\varphi \Leftrightarrow M^-, w' \models \varphi.
    \]
\end{lemma}
\begin{proof}
    We prove this lemma by induction on the complexity of $\varphi \in \mathcal{L}_{IKS}$.
\begin{itemize}
    \item When $\varphi \in Prop$. This can be verified by U-invariance. 
    \item The case of negation and conjunction can be verified by inductive hypothesis. 
    \item When $\varphi = K_c\psi$ such that $c \neq b$.
\begin{itemize}
\item Assume that $M^-|_{a\smalltriangleright b}, w \models K_c\psi$, it means $R_c[w]\subseteq ||\psi||^{a\smalltriangleright b}$. Let $w'R_cu'$ for some $u'$. By Zag$^-$, there is a $u \in M^-|_{a\smalltriangleright b}$ such that $wR_cu$ and $M^-|_{a\smalltriangleright b}, u \overset{a\smalltriangleright b}{\leftrightarroweq} M^-, u'$. From $R_c[w]\subseteq ||\psi||^{a\smalltriangleright b}$ it infers that $u \in ||\psi||^{a\smalltriangleright b}$. By inductive hypothesis and $u \in ||\psi||^{a\smalltriangleright b}$, $M^-, u' \models \psi$. Now we conclude that $M^-, u' \models K_c\psi$.

\item Assume that $M^-, w' \models K_c\psi$ and $wR_cu$. By Zig$^-$ and $wR_cu$, there is a $u'$ such that $w'R_cu'$ and $M^-|_{a\smalltriangleright b}, u \overset{a\smalltriangleright b}{\leftrightarroweq} M^-, u'$. From $M^-, u' \models K_c\psi$ and $w'R_cu'$, it follows $M^-, u' \models \psi$. Again, by inductive hypothesis, we have $M^-|_{a\smalltriangleright b}, u \models \psi$. So it leads to $M^-|_{a\smalltriangleright b}, w\models K_c\psi$. 
\end{itemize}

\item When $\varphi = K_b\psi$.
\begin{itemize}
    \item Assume that $M^-|_{a\smalltriangleright b}, w \models K_b\psi$, it means $R^a_b[w]\subseteq ||\psi||^{M^-|_{a\smalltriangleright b}}$. We need to show $M^-, w' \models K_b\psi$ where $wR_{a\smalltriangleright b}w'$. Let $u' \in M^-$ such that $u' \in R_b[w']$. Given $M^-|_{a\smalltriangleright b}, w \overset{a\smalltriangleright b}{\leftrightarroweq} M^-, w'$ and $u' \in R_b[w']$, due to the Zag$^+$, there is a $u \in M^-|_{a\smalltriangleright b}$ such that $wR^a_bu$ and $M^-|_{a\smalltriangleright b}, u \overset{a\smalltriangleright b}{\leftrightarroweq} M^-, u'$. From $wR^a_bu$ it leads to $u \in ||\psi||^{a\smalltriangleright b}$. Now, by inductive hypothesis, it follows $M^-|_{a\smalltriangleright b}, u \models\varphi \Leftrightarrow M^-, u' \models \varphi$. So $u\in ||\psi||^{M^-}$, which is desired. 

\item Assume that $M^-, w' \models K_b\psi$, which means $R_b[w'] \subseteq ||\psi||^{M^-}$. We want to show $R^a_b[w]\subseteq ||\psi||^{M^-|_{a\smalltriangleright b}}$. Let $u\in R^a_b[w]$. Given $M^-|_{a\smalltriangleright b}, w \overset{a\smalltriangleright b}{\leftrightarroweq} M^-, w'$ and $u\in R^a_b[w]$, due to Zig$^+$, there is a $u' \in M^-$ such that $w'R_bu'$ and $M^-|_{a\smalltriangleright b}, u \overset{a\smalltriangleright b}{\leftrightarroweq} M^-, u'$. From $w'R_bu'$ it follows $u' \in ||\psi||^{M^-}$. According to inductive hypothesis, we have $M^-|_{a\smalltriangleright b}, u \models\varphi \Leftrightarrow M^-, u' \models \varphi$. Hence $u' \in ||\psi||^{M^-|_{a\smalltriangleright b}}$, which is desired.
\end{itemize}

\item When $\varphi = [c\smalltriangleright d]\psi$. We aim to $(M^-|_{a\smalltriangleright b})|_{c\smalltriangleright d}, w \overset{c\smalltriangleright d}{\leftrightarroweq} M^-|_{c\smalltriangleright d}, w'$.
\begin{enumerate}
    \item (Invariance) From the invariance of $M^-|_{a\smalltriangleright b}, w \overset{a\smalltriangleright b}{\leftrightarroweq} M^-, w'$, it leads to $p \in V^{M^-|_{a\smalltriangleright b}}(w) $ if and only if $ p \in V^{M^-}(w')$. By the definition of $(c\smalltriangleright d)$-update, it infers $p \in V^{(M^-|_{a\smalltriangleright b})|_{c\smalltriangleright d}}(w) $ if and only if $ p \in V^{M^-|_{c\smalltriangleright d}}(w')$.
    
    \item (Zig$^-$) Let $w R_e u$ for some $u\in (M^-|_{a\smalltriangleright b})|_{c\smalltriangleright d}$, where $e\neq d$. Because of the definition of update, the unchanged $e$-links, it therefore $w R_e u$ for $w,u\in M^-|_{a\smalltriangleright b}$. Due to the Zig$^-$ of $M^-|_{a\smalltriangleright b}, w \overset{a\smalltriangleright b}{\leftrightarroweq} M^-, w'$, there is a $u'$ such that $w'R_eu'$ and $M^-|_{a\smalltriangleright b},u \overset{a\smalltriangleright b}{\leftrightarroweq} M^-, u'$. By the inductive hypothesis, we can conclude with $(M^-|_{a\smalltriangleright b})|_{c\smalltriangleright d},u \overset{c\smalltriangleright d}{\leftrightarroweq} M^-|_{c\smalltriangleright d}, u'$. 
    
    \item (Zig$^+$) Let $wR^c_d u$ for some $u\in (M^-|_{a\smalltriangleright b})|_{c\smalltriangleright d}$. So $wR_du$ where $u\in M^-|_{a\smalltriangleright b}$. If $d \neq b$, then, similar to the previous application of Zig$^-$, there is a $u'$ such that $w'R_du'$ and $(M^-|_{a\smalltriangleright b})|_{c\smalltriangleright d},u \overset{c\smalltriangleright d}{\leftrightarroweq} M^-|_{c\smalltriangleright d}, u'$. If $d = b$, then it must be $wR^a_bu$ as $w,u \in M^-|_{a\smalltriangleright b}$. By applying Zig$^+$, similarly, it concludes that there is a $u'$ such that $w'R_du'$ and $(M^-|_{a\smalltriangleright b})|_{c\smalltriangleright d},u \overset{c\smalltriangleright d}{\leftrightarroweq} M^-|_{c\smalltriangleright d}, u'$. 
    
    \item (Zag$^-$) Let $w'R_eu'$ for some $u' \in (M^-|_{a\smalltriangleright b})|_{c\smalltriangleright d}$ where $e \neq d$. Again, due to the definition of update, the unchanged $e$-links, it infers $w'R_eu'$ where $w',u' \in M^-|_{a\smalltriangleright b}$. Because of the Zag$^-$ of $M^-|_{a\smalltriangleright b}, w \overset{a\smalltriangleright b}{\leftrightarroweq} M^-, w'$, there is a $u \in M^-|_{a\smalltriangleright b}$ such that $wR_eu$ and $M^-|_{a\smalltriangleright b},u \overset{a\smalltriangleright b}{\leftrightarroweq} M^-, u'$. Again, due to the inductive hypothesis, we can conclude with $(M^-|_{a\smalltriangleright b})|_{c\smalltriangleright d},u \overset{c\smalltriangleright d}{\leftrightarroweq} M^-|_{c\smalltriangleright d}, u'$. 
    
    \item (Zag$^+$) Let $w'R_du'$ for some $u' \in (M^-|_{a\smalltriangleright b})|_{c\smalltriangleright d}$. Due to the definition of update, the removal of the $d$-links, it infers $w'R_du'$ for some $u' \in M^-|_{a\smalltriangleright b}$. When $d\neq b$, similar to the application of Zag$^-$ in the previous case, there is a $u$ such that $wR_du$ in $M^-|_{a\smalltriangleright b}$ and $(M^-|_{a\smalltriangleright b})|_{c\smalltriangleright d},u \overset{c\smalltriangleright d}{\leftrightarroweq} M^-|_{c\smalltriangleright d}, u'$. We need to show $w \equiv_{c:w}u$. From $w'R_du'$ in $(M^-|_{a\smalltriangleright b})|_{c\smalltriangleright d}$ it infers $w' \equiv_{c:w'}u'$. From the invariance of the $K$-modality between $w$ and $w'$ (the previous case) and the inductive hypothesis between $u$ and $u'$, it then leads to $w \equiv_{c:w}u$. So we have $wR^c_du$. When $d = b$, by Zag$^+$ and inductive hypothesis, there is a $u$ such that $wR^a_bu$ in $M^-|_{a\smalltriangleright b}$ and $(M^-|_{a\smalltriangleright b})|_{c\smalltriangleright d},u \overset{c\smalltriangleright d}{\leftrightarroweq} M^-|_{c\smalltriangleright d}, u'$. Similar to the case when $d \neq b$, it has $wR^c_d u$. 
\end{enumerate}
By using $(M^-|_{a\smalltriangleright b})|_{c\smalltriangleright d}, w \overset{c\smalltriangleright d}{\leftrightarroweq} M^-|_{c\smalltriangleright d}, w'$ and the inductive hypothesis, the case of $[c\smalltriangleright d]$ is easily verified. 
\end{itemize}
\end{proof}

\begin{theorem} \label{thm:exeq}
    For any $\varphi \in \mathcal{L}_{IKS}$ in {\bf IKS} and any normal extended model $M$:
    \[
    M,s \models^{\smalltriangleright} \varphi \Leftrightarrow M^-,s \models \varphi.
    \]
\end{theorem}
\begin{proof}
    We only need to discuss the case of $[a\smalltriangleright b]\varphi$.
\begin{itemize}
    \item Let $M,w \models^{\smalltriangleright}[a\smalltriangleright b]\varphi$. So $\forall u \in R^a_b[w]:M, u\models^{\smalltriangleright}\varphi$. By U-Functionality, there is a $u \in R_{a\smalltriangleright b}[w]$. So $M, u\models^{\smalltriangleright}\varphi$. From inductive hypothesis, we know $M^-, u\models\varphi$. From $u \in R_{a\smalltriangleright b}[w]$ and Lemma~\ref{lem:inv}, we know that $M^-|_{a\smalltriangleright b}, w \overset{a\smalltriangleright b}{\leftrightarroweq} M^-, u$. Since {\bf IKS}-formulas are invariant under bisimulation, that is the Lemma~\ref{lem:eq}, we have $M^-|_{a\smalltriangleright b}, w \models \varphi \Leftrightarrow M^-, u \models \varphi$. So we have $M^-, w \models [a\smalltriangleright b]\varphi$, which is desired. 

    \item Suppose $M,w \not\models^{\smalltriangleright}[a\smalltriangleright b]\varphi$. So there is a $u\in R_{a\smalltriangleright b}[w]$ such that $M, u \not\models^{\smalltriangleright}\varphi$. By inductive hypothesis, $M^-, u \not\models\varphi$. We want to show $M^-, w \not\models[a\smalltriangleright b]\varphi$, which is $M^-|_{a\smalltriangleright b}, w \not\models\varphi$. Due to $u\in R_{a\smalltriangleright b}[w]$ and Lemma~\ref{lem:inv}, we have $M^-|_{a\smalltriangleright b}, w \overset{a\smalltriangleright b}{\leftrightarroweq} M^-, u$. So $M^-|_{a\smalltriangleright b}, w \not\models \varphi$ from $M^-, u \not\models\varphi$ and the invariance of bisimulation, the Lemma~\ref{lem:eq}, again.    
    
\end{itemize}
\end{proof}

\begin{definition}[$\smalltriangleright$-Canonical Model] Given the system {\bf IKS}, we construct the $\smalltriangleright$-canonical model $M^{\smalltriangleright} = \langle W^{\smalltriangleright}, \{R_a^{\smalltriangleright}\}_{a \in \agt}, \{R^{\smalltriangleright}_{a:b}\}_{a,b\in\agt},V^{\smalltriangleright}\rangle$ as follows:
\begin{itemize}
    \item $\langle W^{\smalltriangleright}, \{R_a^{\smalltriangleright}\}_{a \in \agt},V^{\smalltriangleright}\rangle$ is a standard canonical model for {\bf IKS}.
    \item $R^{\smalltriangleright}_{a:b} = \{(\Gamma, \Delta) \subseteq W^{\smalltriangleright} \times W^{\smalltriangleright} 
    \mid  \forall \varphi \in \mathcal{L}_{IKS}:[a\smalltriangleright b]\varphi \in \Gamma \Rightarrow \varphi \in \Delta\}.$
\end{itemize}  
\end{definition}
So $(M^{\smalltriangleright})^- = M^C$. As in the standard canonical method~\cite{blackburn2001modal,wang2013axiomatizations}, it is obvious that $\Gamma R^{\smalltriangleright}_{a:b} \Delta$ if and only if $\varphi \in \Delta \Rightarrow \langle a\smalltriangleright b\rangle\varphi\in \Gamma$, where $\langle a\smalltriangleright b\rangle\varphi := \neg[a\smalltriangleright b]\neg\varphi$.


\begin{lemma}[Truth Lemma w.r.t. $\models^{\smalltriangleright}$] \label{lem:truth_tag}
    For any $\varphi \in \mathcal{L}_{IKS}$:
    \[
    M^{\smalltriangleright}, \Delta \models^{\smalltriangleright} \varphi \Leftrightarrow \varphi \in \Delta
    \]
\end{lemma}
\begin{proof}
    All cases of $\varphi$ are standard, however, we will still present the proof of $\varphi = [a\smalltriangleright b]\psi$ here.
\begin{itemize}
    \item Suppose $[a\smalltriangleright b]\psi \in \Delta$. We want to show $M^{\smalltriangleright}, \Delta \models^{\smalltriangleright} [a\smalltriangleright b]\psi$, which is $\forall \Gamma \in R^{\smalltriangleright}_{a:b}[\Delta]: M^{\smalltriangleright}, \Gamma \models^{\smalltriangleright} \psi$. Let $\Gamma \in R^{\smalltriangleright}_{a:b}[\Delta]$. So $\psi \in \Gamma$ by the definition of $R^{\smalltriangleright}_{a:b}$ and $[a\smalltriangleright b]\psi \in \Delta$. By inductive hypothesis, $M^{\smalltriangleright}, \Gamma \models^{\smalltriangleright} \psi$. 
    \item Suppose $\neg[a\smalltriangleright b]\psi \in \Delta$. We want to find a $\Gamma$ such that $\Gamma \in R^{\smalltriangleright}_{a:b}[\Delta]$ and $\Gamma \not\in ||\psi||^{M^{\smalltriangleright}}$. We enumerate the formulas in the countable set $\mathcal{L}_{IKS}$ as $\varphi_1,\dots,\varphi_n,\dots$ and construct $\Gamma_i$ for $i\in\mathbb{N}$ as follows: 
    \begin{align*}
        \Gamma_0 & = \{\neg\psi\} ,\\
        \Gamma_{i+1} & = 
        \begin{cases}
      \Gamma_i \cup \{\varphi_i \mid [a\smalltriangleright b]\varphi_i \in \Delta\}& \text{if this set is {\bf IKS}-consistent; }\\
      \Gamma_i & \text{else. }
    \end{cases}
    \end{align*}
\begin{enumerate}
    \item $\Gamma_i$ ($i\in \mathbb{N}$) is consistent. Otherwise, there are $\psi_1,\dots,\psi_n \in \Gamma_i$ such that $\vdash \neg\psi \to \neg\bigwedge_{1\leq i \leq n}\psi_i$. This is $\vdash \bigwedge_{1\leq i \leq n}\psi_i \to \psi$. By applying the rule {\sf RK} of $[a\smalltriangleright b]$ modality, it infers $\vdash \bigwedge_{1 \leq i \leq n}[a\smalltriangleright b]\psi_i \to [a\smalltriangleright b]\psi$. From $\neg[a\smalltriangleright b]\psi \in \Delta$ it infers $\neg \bigwedge_{1 \leq i \leq n}[a\smalltriangleright b]\psi_i \in\Delta$, which contradicts to $[a\smalltriangleright b]\psi \in \Delta$ and $\Delta$ is in $MCS$. 
    \item We define $\Gamma$ as $\bigcup_{i \in \mathbb{N}}\Gamma_i$. So $\Gamma \in MCS$ as the standard proof. 
    \item $\Gamma \in R^{\smalltriangleright}_{a:b}[\Delta]$ and $\neg\psi \in \Gamma$ can be guaranteed by the construction of $\Gamma$. 
\end{enumerate}
\end{itemize}
\end{proof}

\begin{proposition}[U-Functionality] \label{prop:c1}
    Given any $\Delta \in M^{\smalltriangleright}$, there is a unique $(a\smalltriangleright b)$-successor.
\end{proposition}
\begin{proof} There are two steps for this proof. 
    \begin{itemize}
    \item The existence of a $(a\smalltriangleright b)$-successor. We enumerate the fomulas of the counterble set $\mathcal{L}_{IKS}$ as $\varphi_1, \dots, \varphi_n,\dots$ and construct $\Gamma_i$ for $i\in \mathbb{N}$ as follows. 
\begin{align*}
        \Gamma_0 & = \{\psi \in\mathcal{L}_{IKS} \mid [a\smalltriangleright b]\psi \in \Delta\} ,\\
        \Gamma_{i+1} & = 
        \begin{cases}
      \Gamma_i \cup \{\varphi_i \in \mathcal{L}_{IKS}\}& \text{if this set is {\bf IKS}-consistent; }\\
      \Gamma_i & \text{else. }
    \end{cases}
    \end{align*}
The set $\Gamma_0$ is consistent. If not, there are $[a\smalltriangleright b]\psi_0, [a\smalltriangleright b]\psi_1, \dots, [a\smalltriangleright b]\psi_m \in \Delta$ such that $\vdash \bigwedge_{1\leq i \leq m}\psi_i \to \neg \psi_0$. By the {\sf RK} rule of $[a\smalltriangleright b]$, it leads to $\vdash \bigwedge_{1\leq i \leq m}[a\smalltriangleright b]\psi_i \to [a\smalltriangleright b]\neg \psi_0$. Notice that $[a\smalltriangleright b]\psi_1, \dots, [a\smalltriangleright b]\psi_m \in \Delta$, it then infers $[a\smalltriangleright b]\neg \psi_0 \in \Delta$. According to the {\sf D} axiom, it then has $\neg[a\smalltriangleright b]\psi_0 \in \Delta$, which leads to a contradiction to the assumption $[a\smalltriangleright b]\psi_0 \in \Delta$.  

    \item If $\Gamma, \Gamma'$ are both $\Delta$'s $(a\smalltriangleright b)$-successors, we need to show that they are the same set of formulas. If not, there is a $\varphi \in \Gamma$ and $\neg\varphi \in \Gamma'$. From $\Gamma' \in R^{\smalltriangleright}_{a:b}[\Delta]$, we have $\neg[a\smalltriangleright b]\varphi \in \Delta$. Because of the {\sf Rev} axiom, we then have $[a\smalltriangleright b]\neg\varphi \in \Delta$. Taking together with $\Gamma \in R^{\smalltriangleright}_{a:b}[\Delta]$, it infers $\neg\varphi \in \Gamma$, which leads to a contradiction to $\varphi \in \Gamma$.   
    \end{itemize}
\end{proof}

\begin{proposition}[U-Invariance] \label{prop:c2}
If $\Gamma \in R^{\smalltriangleright}_{a:b}[\Delta]$ then $\forall p \in Prop: p \in \Delta \Leftrightarrow p \in \Gamma$.
\end{proposition}
\begin{proof}
This proposition is guaranteed by the definition of $R^{\smalltriangleright}_{a:b}$ and the {\sf Inv} axiom. 
\end{proof}

\begin{proposition}[U-Zig$^-$] \label{prop:c3}
If $\Delta R^{\smalltriangleright}_a\Gamma$, $\Delta R^{\smalltriangleright}_{b:c} \Delta'$, and $\Gamma R^{\smalltriangleright}_{b:c} \Gamma'$, then $\Delta' R^{\smalltriangleright}_a\Gamma'$.     
\end{proposition}
\begin{proof}
We need to show: $\forall \varphi \in\mathcal{L}_{IKS}: K_a\varphi \in \Delta' \Rightarrow \varphi \in \Gamma'$. Suppose $K_a\varphi \in \Delta'$. So $ \langle b \smalltriangleright c\rangle  K_a\varphi \in \Delta$ from $\Delta R^{\smalltriangleright}_{b:c} \Delta'$. By the {\sf Rev} axiom, we have $[b\smalltriangleright c]K_a\varphi \in \Delta$. Notice that $a\neq c$. By the {\sf Int$^-$} axiom it infers $K_a[b\smalltriangleright c]\varphi \in \Delta$. By applying $\Delta R^{\smalltriangleright}_a\Gamma$ and $\Gamma R^{\smalltriangleright}_{b:c} \Gamma'$, it has $\varphi \in \Gamma'$.  
\end{proof}

\begin{proposition}[U-Zig$^+$] \label{prop:c5}
If $\Delta R^{b\smalltriangleright}_c\Gamma$, $\Delta R^{\smalltriangleright}_{b:c} \Delta'$, and $\Gamma R^{\smalltriangleright}_{b:c} \Gamma'$, then $\Delta' R^{\smalltriangleright}_c\Gamma'$.    
\end{proposition}
\begin{proof}
We want to prove that $K_c\varphi\in\Delta' \Rightarrow \varphi\in \Gamma'$. Let $K_c\varphi\in\Delta'$. So $ \langle b \smalltriangleright c\rangle  K_c\varphi \in \Delta$ from $\Delta R^{\smalltriangleright}_{b:c} \Delta'$. By the {\sf Rev} axiom, we have $[b\smalltriangleright c]K_c\varphi \in \Delta$. By applying the {\sf Int$^+$} axiom, there is a $\psi \in \mathcal{L}_{IKS}$ such that $K_c[b\smalltriangleright c](\psi \to \varphi) \in \Delta$ and $K_b[b\smalltriangleright c]\psi\in \Delta$. So, from $\Delta R^{b\smalltriangleright}_c\Gamma$ and $K_c[b\smalltriangleright c](\psi \to \varphi) \in \Delta$, it follows $[b\smalltriangleright c](\psi \to \varphi) \in \Gamma$. From $\Gamma R^{\smalltriangleright}_{b:c}\Gamma'$ it leads to $\psi \to \varphi \in \Gamma'$. On the other hand, from $\Delta R^{b\smalltriangleright}_c\Gamma$ it indicates $\Delta \equiv_{b:\Delta}\Gamma$, which means that $\forall \varphi \in K_b\Delta: \varphi \in \Delta \Leftrightarrow \varphi \in \Gamma$. We already have $K_b[b\smalltriangleright c]\psi\in \Delta$. From the {\sf T} axiom of the $K_b$ modality it is the case that $[b\smalltriangleright c]\psi \in \Delta$. So $[b\smalltriangleright c]\psi \in \Gamma$. From $\Gamma R^{\smalltriangleright}_{b:c}\Gamma'$ it infers $\psi \in \Gamma'$. Now we conclude with $\varphi \in \Gamma'$.  
\end{proof}

\begin{proposition}[U-Zag$^-$] \label{prop:c4}
If $\Delta' R^{\smalltriangleright}_a\Gamma'$ and $\Delta R^{\smalltriangleright}_{b:c} \Delta'$, then there is a $\Gamma \in MCS$ such that $\Gamma R^{\smalltriangleright}_{b:c} \Gamma'$ and $\Delta R^{\smalltriangleright}_a\Gamma$.    
\end{proposition}
\begin{proof}
    Notice that $a\neq c$. We enumerate $\varphi_1,\dots,\varphi_n,\dots$ in $\mathcal{L}_{IKS}$ and construct $\Gamma_i$ ($i\in \mathbb{N}$) as follows:
    \begin{align*}
        \Gamma_0 & = \{\varphi \in \mathcal{L}_{IKS} \mid K_a\varphi \in \Delta\} \cup \{\langle b\smalltriangleright c\rangle\varphi \in \mathcal{L}_{IKS} \mid \varphi \in \Gamma' \} ,\\
        \Gamma_{i+1} & = 
        \begin{cases}
      \Gamma_i \cup \{\varphi_i \in \mathcal{L}_{IKS}\}& \text{if this set is {\bf IKS}-consistent; }\\
      \Gamma_i & \text{else. }
    \end{cases}
    \end{align*}
    \begin{itemize}
        \item $\Gamma_0$ is consistent. If not, there is $K_a\varphi_1,\dots,K_a\varphi_n\in\Delta$ and $\psi_1, \dots, \psi_m\in\Gamma'$ such that $\vdash \bigwedge_{1\leq i \leq n}\varphi_i \wedge \bigwedge_{1\leq j \leq m}\langle b\smalltriangleright c\rangle\psi_j \to \bot$. This means $\vdash \bigwedge_{1\leq i \leq n}\varphi_i \to \neg \bigwedge_{1\leq j \leq m}\langle b\smalltriangleright c\rangle\psi_j$, which is $\vdash \bigwedge_{1\leq i \leq n}\varphi_i \to  \bigvee_{1\leq j \leq m}[ b\smalltriangleright c]\neg\psi_j$. By using the {\sf RM$_{\smalltriangleright}$} rule, it is $\vdash \bigwedge_{1\leq i \leq n}K_a\varphi_i \to  \bigvee_{1\leq j \leq m}[ b\smalltriangleright c]K_a\neg\psi_j$. From $K_a\varphi_i\in \Delta$ for any $i$, it has $\bigwedge_{1\leq i \leq n}K_a\varphi_i \in \Delta$. So $\bigvee_{1\leq j \leq m}[ b\smalltriangleright c]K_a\neg\psi_j \in \Delta$. Suppose $[ b\smalltriangleright c]K_a\neg\psi_k \in \Delta$ where $\psi_k\in \Gamma'$. From $\Delta R^{\smalltriangleright}_{b:c} \Delta'$, it follows $K_a \neg\psi_k \in \Delta'$. Because $\Delta' R^{\smalltriangleright}_a\Gamma'$, it leads to $\neg\psi_k\in\Gamma'$, which is a contradiction.
           
        \item $\Gamma_i$ for each $i \in \mathbb{N}$ is consistent. The proof is similar to the case of $\Gamma_0$.
        \item We define $\Gamma = \bigcup_{i \in \mathbb{N}}\Gamma_i$. So $\Gamma \in MCS$. 
        \item The construction of $\Gamma$ ensures that $\Gamma R^{\smalltriangleright}_{b:c} \Gamma'$ and $\Delta R^{\smalltriangleright}_a\Gamma$. 
    \end{itemize}
\end{proof}

\begin{lemma} \label{lem:updateK}
If $\Delta R^{\smalltriangleright}_{b:c}\Delta'$ then $K_b\varphi \in \Delta' \Rightarrow [b\smalltriangleright c]K_c\varphi \in \Delta$.    
\end{lemma}
\begin{proof}
When $\Delta R^{\smalltriangleright}_{b:c}\Delta'$, it is $\forall\varphi \in \mathcal{L}_{IKS}: \varphi \in \Delta' \Rightarrow [b\smalltriangleright c]\varphi \in \Delta$ by using the {\sf Rev} axiom. From the left-to-right side of the {\sf Rep} axiom, it leads to $\forall\varphi \in \mathcal{L}_{IKS}: \varphi \in \Delta' \Rightarrow [b\smalltriangleright c]\varphi \in \Delta'$. From $\Delta R^{\smalltriangleright}_{b:c}\Delta'$, it has $\varphi \to [b\smalltriangleright c]\varphi$ valid. So $K_b\varphi \to [b\smalltriangleright c]K_c\varphi$ valid from the {\sf Inc$_{\smalltriangleright}$} rule. If $K_b\varphi \in \Delta'$ then $[b\smalltriangleright c]K_c\varphi \in \Delta'$. From $\Delta R^{\smalltriangleright}_{b:c}\Delta'$ and the {\sf Rev} axiom again, $[b \smalltriangleright c][b\smalltriangleright c]K_c\varphi \in \Delta$. From the right-to-left side of the {\sf Rep} axiom, it leads to $[b\smalltriangleright c]K_c\varphi \in \Delta$.        
\end{proof}

\begin{lemma} \label{lem:stay}
If $\Delta R^{\smalltriangleright}_{b:c}\Delta'$ and $\Delta'R^{\smalltriangleright}_c\Gamma'$ then $\Delta'\equiv_{b:\Delta'}\Gamma'$.
\end{lemma}
\begin{proof}
We need to show: $\forall \varphi \in K_b\Delta': \varphi \in \Delta' \Leftrightarrow \varphi \in \Gamma'$. Given $\varphi \in K_b\Delta'$:
\begin{itemize}
\item If $\varphi \in \Delta'$. From $\varphi \in K_b\Delta'$, it has $K_b\varphi\in\Delta'$. So $[b\smalltriangleright c]K_c\varphi \in \Delta$ from Lemma~\ref{lem:updateK}. From $\Delta R^{\smalltriangleright}_{b:c}\Delta'$ it infers $K_c\varphi \in \Delta'$. From $\Delta'R^{\smalltriangleright}_c\Gamma'$ it leads to $\varphi \in \Gamma'$. 
\item If $\varphi \in \Gamma'$. From $\varphi \in K_b\Delta'$, it has $K_b\varphi\in\Delta'$. From the {\sf T} axiom of the $K_b$ modality, it infers $\varphi \in \Delta'$, which is desired.  
\end{itemize}  
\end{proof}

\begin{proposition}[U-Zag$^+$] \label{prop:c6}
If $\Delta' R^{\smalltriangleright}_c\Gamma'$ and $\Delta R^{\smalltriangleright}_{b:c} \Delta'$, then there is a $\Gamma \in MCS$ such that $\Gamma R^{\smalltriangleright}_{b:c} \Gamma'$ and $\Delta R^{b\smalltriangleright}_c\Gamma$.    
\end{proposition}
\begin{proof}
We enumerate $\varphi_1,\dots,\varphi_n,\dots$ in $\mathcal{L}_{IKS}$ and construct $\Gamma_i$ ($i\in \mathbb{N}$) as follows:
    \begin{align*}
        \Gamma^0 & = \{\varphi  \mid K_c\varphi \in \Delta\} \cup \{\langle b\smalltriangleright c\rangle\varphi \mid \varphi \in \Gamma' \} ,\\
        \Gamma_0 & = \Gamma^0 \cup \{\varphi \mid K_b\varphi \in \Delta'\}\\
        \Gamma_{i+1} & = 
        \begin{cases}
      \Gamma_i \cup \{\varphi_i \in \mathcal{L}_{IKS}\}& \text{if this set is {\bf IKS}-consistent; }\\
      \Gamma_i & \text{else. }
    \end{cases}
    \end{align*}
Similar to the proof in Proposition~\ref{prop:c4}, $\Gamma^0$ is consistent. We want to show: $\Gamma_0$ is consistent. Otherwise, $\exists K_b\chi_1, \dots, K_b\chi_k \in \Delta$, $\exists K_c\varphi_1,\dots,K_c\varphi_n \in \Delta$, and $\exists \psi_1,\dots, \psi_m\in \Gamma'$ such that 
\[
\vdash \bigwedge_{1\leq l\leq k}\chi_l \to \neg(\bigwedge_{1\leq i\leq n}\varphi_i\wedge\bigwedge_{1\leq j\leq m}\langle b\smalltriangleright c\rangle \psi_j).
\]
By the {\sf NEC} rule of the $K_b$ modality, it leads to
\[
\vdash K_b\bigwedge_{1\leq l\leq k}\chi_l \to K_b\neg(\bigwedge_{1\leq i\leq n}\varphi_i\wedge\bigwedge_{1\leq j\leq m}\langle b\smalltriangleright c\rangle \psi_j).
\]
From $K_b\chi_1, \dots, K_b\chi_k \in \Delta$, it is $K_b\neg(\bigwedge_{1\leq i\leq n}\varphi_i\wedge\bigwedge_{1\leq j\leq m}\langle b\smalltriangleright c\rangle \psi_j).$ By applying either the {\sf Int$_+$} or the {\sf Int$^-$} axioms, it infers 
\[
\bigwedge_{1\leq i\leq n}K_b\varphi_i \to \bigvee_{1\leq j\leq m}[ b\smalltriangleright c]K_b\neg \psi_j \in \Delta.
\]
From $\Delta R^{\smalltriangleright}_{b:c} \Delta'$ it follows
\[
\bigwedge_{1\leq i\leq n}K_b\varphi_i \in \Delta \Rightarrow \bigvee_{1\leq j\leq m}K_b\neg \psi_j \in \Delta'.
\]
So, there is a $\psi_k \in \Gamma'$ such that 
\[
\bigwedge_{1\leq i\leq n}K_b\varphi_i \in \Delta \Rightarrow K_b\neg \psi_k \in \Delta'.
\]
From $\Delta' R^{\smalltriangleright}_c\Gamma'$ and $\Delta R^{\smalltriangleright}_{b:c} \Delta'$, and Lemma~\ref{lem:stay}, we know that $\Delta'\equiv_{b:\Delta'}\Gamma'$. So, when $K_b\neg \psi_k \in \Delta'$, due to the {\sf T} axiom of the $K_b$ modality, it is the case that
\[
\bigwedge_{1\leq i\leq n}K_b\varphi_i \in \Delta \Rightarrow \neg \psi_k \in \Gamma'.
\]
Notice that $\psi_k \in \Gamma'$, there is a $\varphi_i$ such that $\neg K_b\varphi_i \in \Delta$. However, it is in conflict with the assumption. So $\Gamma_0$ is consistent too. The consistency of all other $\Gamma_i$ can be verified easily. 

We define $\Gamma = \bigcup_{i \in \mathbb{N}}\Gamma_i$. So $\Gamma \in MCS$. The construction of $\Gamma$ ensures that $\Gamma R^{\smalltriangleright}_{b:c} \Gamma'$ and $\Delta R^{\smalltriangleright}_c\Gamma$. It also ensures $\Delta R^{\smalltriangleright}_b\Gamma$. From the proof of Lemma~\ref{lem:u4}, we know that $R^{\smalltriangleright}_b \subseteq \equiv_{b:\Delta}$. So $\Delta\equiv_{b:\Delta}\Gamma$.       
\end{proof}

\begin{proposition}\label{prop:cc}
    $M^{\smalltriangleright}$ is a normal extended model of $M^C$.
\end{proposition}
\begin{proof}
    This proposition can be verified by Proposition~\ref{prop:c1}, Proposition~\ref{prop:c2},  Proposition~\ref{prop:c3}, Proposition~\ref{prop:c5}, Proposition~\ref{prop:c4}, and Proposition~\ref{prop:c6}. 
\end{proof}

\begin{lemma}[Truth Lemma]
    For any $\varphi \in \mathcal{L}_{IKS}$ and a point $w \in M^c$:
    \[
    \varphi \in w \Leftrightarrow M^C, w \models \varphi
    \]
\end{lemma}
\begin{proof}
    This lemma can be verified by using Theorem~\ref{thm:exeq} and Lemma~\ref{lem:truth_tag} and by $(M^{\smalltriangleright})^- = M^C$ from Proposition~\ref{prop:cc}.
\end{proof}


\subsection{Other Systems}

According to the languages given in Section~\ref{sec:express}, we can continue axiomatizing the logical systems of knowledge sharing as follows. The system {\bf IK} is the S5 proof system of individual knowledge~\cite{van2007dynamic}. The system {\bf AKS} contains all axioms and rules both in {\bf AK} and {\bf IKS}, along with these two interaction axioms:
\begin{itemize}
\item $[a\smalltriangleright b]K_b \varphi\leftrightarrow K^a_b[a\smalltriangleright b]\varphi$ and
\item $[a\smalltriangleright b]K^a_b \varphi\leftrightarrow K^a_b[a\smalltriangleright b]\varphi$.
\end{itemize}
The system {\bf DK} contains all axioms and rules in the classical proof system for distributed knowledge~\cite{van2007dynamic} and in the system {\bf AK}. The system {\bf DKS} contains all axioms and rules in {\bf DK} and in {\bf ASK} together with this axiom:
\begin{itemize}
    \item $[a\smalltriangleright b]D_G\varphi \leftrightarrow D_G[a\smalltriangleright b]\varphi$.
\end{itemize}

\begin{proposition}
$[a\smalltriangleright b] K_b\varphi \to D_{\{a,b\}}[a\smalltriangleright b]\varphi$ is derivable in {\bf DKS}.
\end{proposition}
\begin{proof}
\begin{itemize}
\item It can be verified by {\sf Int$^+$} axiom with the interaction axiom of $D_G$ modality. 
\end{itemize}
\end{proof}

The system {\bf DKR} extends {\bf DKS} with the axioms and rules in the system of resolved information~\cite{aagotnes2017resolving} as well as this axiom:
\begin{itemize}
    \item ({\sf Int$^{R}$}) $R_GK_a\varphi  \to R^i_GK_a\varphi$. 
\end{itemize}
\begin{proposition}
    {\sf Int$^{R}$} is valid.
\end{proposition}
\begin{proof}
\begin{itemize}
    \item Suppose $M, w \models R_G\varphi$. This indicates $ R_a \cap \equiv_{G\setminus\{a\}}\subseteq ||\varphi||^*$ where $a \in G$. Notice that $\bigcap_{i\in G} R_i \subseteq R_a \cap \equiv_{G\setminus\{a\}}$ from the proof of Lemma~\ref{lem:update}.\ref{lem:u4}. So $\bigcap_{i\in G} R_i \subseteq ||\varphi||^*$ together from applying {\sf Boolean}.   
\end{itemize}
\end{proof}

\begin{proposition} 
These are derivable in {\bf DKR}:
\begin{enumerate}
    \item $K_{a_1}\varphi \to R^{a_1}_GE_G\varphi$ where $\varphi \in Prop^+$ 
    \item $R^{a_1}_GE_G\varphi \to R_GE_G\varphi$ where $\varphi \in Prop^+$  
    \item $E_G\varphi \to R_GE_G\varphi$ where $\varphi \in Prop^+$  
    \item $R_GE_G\varphi \to R^i_GE_G\varphi$ where $\varphi \in Prop^+$  
\end{enumerate}
\end{proposition}
\begin{proof}
\begin{enumerate}
    \item The following statements are derived from the axioms and rules of {\sf Sharing, Remain, N$^s$, K$_{\smalltriangleright}$}, and {\sf MP}:
\begin{align*}
    K_{a_1}\varphi \to [a_1\smalltriangleright a_2](K_{a_1}\varphi \wedge K_{a_2}\varphi) \\ \text{ by {\sf Sharing} and {\sf Remain}} \\
    [a_1\smalltriangleright a_2](K_{a_1}\varphi \wedge K_{a_2}\varphi) \to [a_1\smalltriangleright a_2][a_2\smalltriangleright a_3](K_{a_1}\varphi \wedge K_{a_2}\varphi \wedge K_{a_3}\varphi) \\ \text{ by {\sf Sharing, Remain, N$^s$, K$_{\smalltriangleright}$}, and {\sf MP}}\\
    \cdots\\
    [a_1\smalltriangleright a_2]\cdots[a_{n-2}\smalltriangleright a_{n-1}]\bigwedge_{1\leq i \leq n-1}K_{a_i}\varphi \to [a_1\smalltriangleright a_2]\cdots[a_{n-1}\smalltriangleright a_{n}]\bigwedge_{1\leq i \leq n}K_{a_i}\varphi
\end{align*}
So we have $K_{a_1}\varphi \to [a_1\smalltriangleright a_2]\cdots[a_{n-1}\smalltriangleright a_{n}]\bigwedge_{1\leq i \leq n}K_{a_i}\varphi$, which is the desired result $K_{a_1}\varphi \to R_GE_G\varphi$.

\item The following statements are derivable from Proposition~\ref{prop:RG}.\ref{prop:rg1}, {\sf N$_s$, K$_{\smalltriangleright}$}, and {\sf MP}:
\begin{align*}
    E_G\varphi \to R^{a_n}_GE_G\varphi
    &\quad\text{ by Proposition~\ref{prop:RG}.\ref{prop:rg1}}\\
    [a_{n-1}\smalltriangleright a_n](E_G\varphi \to R^{a_n}_GE_G\varphi)
    &\quad\text{ by {\sf N$_s$}}\\
    [a_{n-1}\smalltriangleright a_n]E_G\varphi \to [a_{n-1}\smalltriangleright a_n]R^{a_n}_GE_G\varphi
    &\quad\text{ by {\sf K$_{\smalltriangleright}$} and {\sf MP}}\\
    \dots&\dots\\
    R^{a_1}_GE_G\varphi \to R_GE_G\varphi &\quad\text{by {\sf N$_s$, K$_{\smalltriangleright}$} and {\sf MP}}
\end{align*}

\item Similar to Proposition~\ref{prop:RG}.\ref{prop:rg1}, by {\sf Sharing, Remain, N$_s$, K$_{\smalltriangleright}$}, and {\sf MP}, this formula is derivable.  

\item This can be derived from the {\sf Int$^{R}$} axiom. 
\end{enumerate}
\end{proof}
So we have the following implication in {\bf DKR}:
\[
E_G\varphi \to K_a\varphi \to R^a_GE_G\varphi \to R_GE_G\varphi \to R^i_GE_G\varphi.
\]

\section{Related Work} \label{sec:related}

\subsection{Methodology of Link Deletion}
Our model update, focused on $(a\smalltriangleright b)$-knowledge sharing, selectively eliminates all $b$-links within the model that do not align with agent $a$'s knowledge. In the field of model update and modification, deletion is widely recognized as a core strategy, as evidenced by various methodologies including public announcement~\cite{plaza1989logics}, semantic updates~\cite{veltman1996defaults}, product updates~\cite{baltag2004logics}, upgrades~\cite{VB2007dynamic}, and sabotage updates~\cite{van2005essay} among others. Technical surveys detailing the results of their complexities can referred to~\cite{van2007dynamic,van2011logical,aucher2018modal}. Link deletion, in particular, is essential for defining strategies in information pooling, where each method of update is designed for a specific function: Product updates~\cite{baltag2004logics,van2007dynamic} are agent-oriented, upgrades~\cite{VB2007dynamic} focus on formula-specific modifications, and sabotage methods~\cite{van2005essay,van2011logical} target the removal of arbitrary links without regard to specific expressions~\footnote{A recent proposal introduces a language-specific approach to sabotage updates~\cite{aucher2018modal}. Furthermore, an exploration of updates from a language-specific perspective to those unrelated to language in sabotage modifications was presented in~\cite{van2022modal}. }. In the domain of information pooling, algebraic intersection methods that consolidate all potential information relative to specific agents~\cite{aagotnes2017resolving,van2022pooling,castaneda2023communication,baltag2023learning} stand out. Unlike previous approaches, our method remains agent-specific in the syntax in its deletion of links, but notably underscores the \emph{semantic} role of knowledge formulas in the update process.
 
\subsection{Social Dependence}
Our knowledge pooling model brings to light a dimension of dependence within social interactions, a concept extensively examined in various formal theories. The notion of belief dependence, as introduced by Delgrande et al.~\cite{del1996belief}, focuses on belief change and revision, thereby framing \emph{epistemic} dependence. However, it is formalized within conditional logics employing a static language framework. On a different note, Boella et al.~\cite{boella2007social} have proposed a formal model designed to explain the power dynamics inherent in social dependence, rooted in an agent's reliance on others to achieve their goals. This model of \emph{normative} dependence utilises Castelfranchi's methodology of dependence graphs~\cite{castelfranchi2003micro}, albeit without integrating a logical system.

In contrast, our framework for permissible knowledge pooling introduces a concept of dependence that not only underscores the dynamics of knowledge among individuals but also explores the capacity for maintaining the permissibility of information sharing between agents. This approach sheds light on the complex network of interpersonal dependencies, examined through \emph{both epistemic and normative lenses}. Thus, this approach holds promise for future explorations into understanding social hierarchies and power dynamics.

\subsection{Permission to Know}
Our methodology for permissible knowledge pooling draws on the established logic of permission, as introduced by van der Meyden~\cite{van1996dynamic}, which identifies permissibility across pairs of states concerning the authorization to acquire knowledge. Traditionally, most models have formalized the concept of permission to know by highlighting permissibility in relation to states. Moreover, inspired from Aucher et al.~\cite{aucher2010privacy}, several existing studies view permission to know as contingent upon permission to communicate—postulating that knowledge acquired through communication should not result in a breach of permissible states, as supported by the research of Balbiani et al.~\cite{balbiani2011reasoning}, Li et al. ~\cite{li2022dynamic}, and Van Ditmarsch and Seban~\cite{van2012logical}.

Our approach introduces a different perspective on the interplay between the permission to know and the permission to communicate, emphasising that the former precedes the latter. In essence, our framework suggests that we are granted the permission to know prior to receiving the authorization to communicate. This conceptualization of permissible knowledge pooling is distinctly defined through the lens of permission to know, offering a fresh viewpoint on its relationship with communicative permissions.

\section{Conclusions} \label{sec:con}

This study explores two key aspects of multi-agent communication: knowledge pooling and the permission to share knowledge. Unlike the conventional approach to information pooling, which aggregates arbitrary information to form collective knowledge, our framework selectively integrates only verified and known information during the pooling process. This approach introduces a novel perspective to information pooling, uncovering a unique form of collective knowledge that bridges the gap between individual and distributed knowledge. 

Furthermore, our investigation offers a unique viewpoint on the interaction between epistemic and ethical considerations, successfully establishing a framework that explains how epistemic states together with our normative entitlement to knowledge influence our actions. By examining the relationship between the permission to know and the permission to share knowledge, we provide insightful observations on this interplay.

In our formal framework, we have investigated a range of validities and expressivities related to knowledge pooling, distributed knowledge, and the permission to know, emphasizing their roles within permissible knowledge pooling. We also explore their axiomatizations in addition with the soundness and completeness of systems {\bf AK} for agent-dependent knowledge and {\bf IKS} for dynamic knowledge sharing. The soundness and completeness of other systems are straightforward based on the results of {\bf AK} and {\bf IKS}. We leave the exploration of computational complexity and decidability to future technical research.

Our framework for permissible knowledge pooling introduces various challenges in the formalization of social reasoning within multi-agent interactions. For example, investigating the applicability of our concept of knowledge pooling -- as a form of social dependence -- to interpret social power is of significant interest. This includes examining the correlation between the extent of knowledge sharing and the structure of social hierarchies. Additionally, incorporating the concept of common knowledge into our framework presents an opportunity to explore deeper into the relationship between knowledge pooling and the achievement of social consensus.

\section*{Acknowledgement}
This work is supported by the \emph{Fonds National de la Recherche} of Luxembourg through the project Deontic Logic for Epistemic Rights (OPEN O20/14776480).


\bibliographystyle{aiml}
\bibliography{aiml.bib}

\end{document}

%% file: TabFig/fig_3k.tex
\tikz[every lower node part/.style={black}] {
  \begin{scope}[name prefix = 2-]

  \node[shape=circle,draw=black, label=right:{\footnotesize$p,q,r$},scale=.7] (0) at (0,0) {$s_0$};
  \node[shape=circle,draw=black, label=left:{\footnotesize$p,q,  \neg r$},scale=0.7] (1) at (-2,1) {$s_1$};
  \node[shape=circle,draw=black, label=left:{\footnotesize$\neg p,q,  r$},scale=0.7] (2) at (-2,-1) {$s_2$};
   \node[shape=circle,draw=black, label=right:{\footnotesize$p, \neg  q, \neg r$},scale=0.7] (3) at (2,1) {$s_3$};
  \node[shape=circle,draw=black, label=right:{\footnotesize$\neg  p,\neg   q, \neg r$},scale=0.7] (4) at (2,-1) {$s_4$};

    \draw[green!50!black!80!,dashed] (0) edge node[above, name=a] {\footnotesize$ac$} (1);
\draw[black,dashed] (0) edge node[below, name=a] {\footnotesize$ac$} (2); 
    \draw[green!50!black!80!,dashed](0) edge node[above, name=b] {\footnotesize$bc$} (3);
    \draw[black,dashed](0) edge node[below, name=b] {\footnotesize$bc$} (4);
  \end{scope}
}

%% file: TabFig/fig_ks.tex
\tikz[every lower node part/.style={black}] {
  \begin{scope}[name prefix = 2-]
    \node [circle split,draw, minimum size=1.2cm,scale=0.7] (0) at (0,0) {$s_0$ \nodepart{lower} \footnotesize$p,q,r$};
    \node [circle split,draw, minimum size=1.2cm,scale=0.7] (1) at (-2.3,0) {$s_1$ \nodepart{lower} \footnotesize$p,\neg q, \neg r$};
    \node [circle split,draw, minimum size=1.2cm,scale=0.7] (2) at (2,1) {$s_2$ \nodepart{lower} \footnotesize$p, \neg q,r$};
    \node [circle split,draw, minimum size=1.2cm,scale=0.7] (3) at (2,-1) {$s_3$ \nodepart{lower} \footnotesize$p, q, \neg r$};
    \path[draw, dashed] 
    (0) edge node[above, name=b] {\footnotesize$a$} (1)
    (0) edge node[above, name=b] {\footnotesize$b$} (2) 
    (0) edge node[below, name=b] {\footnotesize$b$} (3);
  \end{scope}
}

%% file: TabFig/fig_dire1.tex
\tikz[every lower node part/.style={black}] {
  \begin{scope}[name prefix = 2-]

  \node[shape=circle,draw=black, label=right:{\footnotesize$p,q,r$},scale=.7] (0) at (0,0) {$s_0$};
  \node[shape=circle,draw=black, label=left:{\footnotesize$p,q,  \neg r$},scale=0.7] (1) at (-1.5,1) {$s_1$};
  \node[shape=circle,draw=black, label=left:{\footnotesize$\neg p,q,  r$},scale=0.7] (2) at (-1.5,-1) {$s_2$};
   \node[shape=circle,draw=black, label=right:{\footnotesize$p, \neg  q, \neg r$},scale=0.7] (3) at (1.5,1) {$s_3$};
  \node[shape=circle,draw=black, label=right:{\footnotesize$\neg  p,\neg   q, \neg r$},scale=0.7] (4) at (1.5,-1) {$s_4$};

     \draw[green!50!black!80!,dashed] (0) edge node[above, name=a] {\footnotesize$ac$} (1);
\draw[black,dashed] (0) edge node[below, name=a] {\footnotesize$ac$} (2); 
    \draw[green!50!black!80!,dashed](0) edge node[above, name=b] {\footnotesize$b$} (3);
    \draw[black,dashed](0) edge node[below, name=b] {\footnotesize$b$} (4);
  \end{scope}
}

%% file: TabFig/fig_dire2.tex
\tikz[every lower node part/.style={black}] {
  \begin{scope}[name prefix = 2-]

  \node[shape=circle,draw=black, label=right:{\footnotesize$p,q,r$},scale=.7] (0) at (0,0) {$s_0$};
  \node[shape=circle,draw=black, label=left:{\footnotesize$p,q,  \neg r$},scale=0.7] (1) at (-1.5,1) {$s_1$};
  \node[shape=circle,draw=black, label=left:{\footnotesize$\neg p,q,  r$},scale=0.7] (2) at (-1.5,-1) {$s_2$};
   \node[shape=circle,draw=black, label=right:{\footnotesize$p, \neg  q, \neg r$},scale=0.7] (3) at (1.5,1) {$s_3$};
  \node[shape=circle,draw=black, label=right:{\footnotesize$\neg  p,\neg   q, \neg r$},scale=0.7] (4) at (1.5,-1) {$s_4$};

     \draw[green!50!black!80!,dashed] (0) edge node[above, name=a] {\footnotesize$a$} (1);
\draw[black,dashed] (0) edge node[below, name=a] {\footnotesize$ac$} (2); 
    \draw[green!50!black!80!,dashed](0) edge node[above, name=b] {\footnotesize$bc$} (3);
    \draw[black,dashed](0) edge node[below, name=b] {\footnotesize$bc$} (4);
  \end{scope}
}

%% file: TabFig/fig_express.tex
\centering
\begin{tikzpicture}[node distance = 2cm, auto]
\node (IK) at (1,4) {$\mathcal{L}_{IK}$};
\node (AK) at (3,4) {$\mathcal{L}_{AK}$};
\node (DK) at (5,4) {$\mathcal{L}_{DK}$};
\node (IKS) at (1,3) {$\mathcal{L}_{IKS}$};
\node (AKS) at (3,3) {$\mathcal{L}_{AKS}$};
\node (DKS) at (5,3) {$\mathcal{L}_{DKS}$};
\node (DKR) at (7,3) {$\mathcal{L}_{DKR}$};

\draw[->] (IK) to  node {} (AK);
\draw[->] (AK) to  node {} (DK);
\draw[->] (AK) to  node {} (AKS);
\draw[->] (IK) to  node {} (IKS);
\draw[->] (IKS) to  node {} (AKS);
\draw[->] (AKS) to  node {} (DKS);
\draw[->] (DK) to  node {} (DKR);
\draw[->] (DK) to  node {} (DKS);
\draw[->] (DKS) to  node {} (DKR);
\end{tikzpicture}